\documentclass[12pt]{iopart}

\usepackage{graphicx}
\usepackage{amssymb}
\usepackage[usenames,dvipsnames]{xcolor}

\newtheorem{Th}{Theorem}

\newtheorem{Prop}{Proposition}
     
\newcommand{\be}{\begin{equation}}
\newcommand{\ee}{\end{equation}} 
\newcommand{\lb}{\label}
\newcommand{\ol}{\overline}

\newcommand{\btau}{{\mbox{\boldmath $\tau$}}}

\newcommand{\const}{({\rm const.})}

\newcommand{\ba}{{\bf a}}

\newcommand{\bF}{{\bf f}}

\newcommand{\bn}{{\bf n}}
\renewcommand{\br}{{\bf r}}
\newcommand{\bu}{{\bf u}}

\newcommand{\bx}{{\bf x}}

\newcommand{\bA}{{\bf A}}
\newcommand{\bB}{{\bf B}}
\newcommand{\bP}{{\bf P}}
\newcommand{\bE}{{\bf E}}

\newcommand{\bJ}{{\bf J}}
\newcommand{\bS}{{\bf S}}

\newcommand{\fltr}{\overline}
\newcommand{\OL}{\overline}
\newcommand{\fB}{\fltr\bB}
\newcommand{\fu}{\fltr\bu}
\newcommand{\fS}{\fltr\bS}
\newcommand{\fJ}{\fltr\bJ}
\newcommand{\fA}{\fltr\bA}
\newcommand{\fomega}{\fltr\bomega}

\newcommand{\bomega}{{\mbox{\boldmath $\omega$}}}
\newcommand{\bepsilon}{{\mbox{\boldmath $\varepsilon$}}}

\newcommand{\grad}{{\mbox{\boldmath $\nabla$}}}
\newcommand{\bdot}{{\mbox{\boldmath $\cdot$}}}

\newcommand{\btimes}{{\mbox{\boldmath $\times$}}}
\newcommand{\bzed}{{\mbox{\boldmath $0$}}}

\begin{document}

\title[Coarse-Grained Incompressible Magnetohydrodynamics]{Coarse-Grained Incompressible Magnetohydrodynamics: analyzing the turbulent cascades}

\author{Hussein Aluie$^{1,2}$}

\address{$^1$Department of Mechanical Engineering,}
\address{$^2$Laboratory for Laser Energetics,\\ ~University of Rochester, Rochester, NY 14627, USA}
\vspace{10pt}
\begin{indented}
\item[]\today
\end{indented}

\begin{abstract}
We formulate a coarse-graining approach to the dynamics of magnetohydrodynamic (MHD) fluids 
at a continuum of length-scales $\ell$. In this methodology, effective equations are derived for the observable 
velocity and magnetic fields spatially-averaged at an arbitrary scale of resolution. 
The microscopic equations for the ``bare'' velocity and 
magnetic fields are ``renormalized'' by coarse-graining to yield macroscopic effective equations that contain 
both a subscale stress and a subscale electromotive force (EMF) generated by nonlinear interaction of eliminated 
fields and plasma motions. Particular attention is given to the effects of these subscale terms on the balances 
of the quadratic invariants of ideal incompressible MHD---energy, cross-helicity and magnetic helicity. 
At large coarse-graining length-scales, the direct dissipation of the invariants by microscopic 
mechanisms (such as molecular viscosity and Spitzer resistivity) is shown to be negligible. The balance 
at large scales is dominated instead by the subscale nonlinear terms, which can transfer invariants 
across scales, and are interpreted in terms of work concepts for energy and in terms of topological 
flux-linkage for the two helicities. An important application of this approach is to MHD turbulence, 
where the coarse-graining length $\ell$ lies in the inertial cascade range. 
We show that in the case of sufficiently rough velocity and/or magnetic fields, 
the nonlinear inter-scale transfer need not vanish and can persist to arbitrarily small scales.
 Although closed expressions are not available for subscale stress 
and subscale  EMF, we derive rigorous upper bounds on the effective dissipation they produce in terms 
of scaling exponents of the velocity and magnetic fields. These bounds provide exact constraints 
on  phenomenological theories of MHD  turbulence in  order to allow the nonlinear cascade of energy and 
cross-helicity.  On the other hand, we prove a very strong version of the Woltjer-Taylor conjecture on 
conservation of magnetic helicity. Our bounds show that forward cascade of magnetic helicity 
to asymptotically small scales is impossible unless 3rd-order moments of either velocity or magnetic field become infinite. \end{abstract}

%
%
%
%
%

\section{Introduction}

All  measurements in  laboratory experiments and in nature have a limited resolution 
in space and time. This is particularly true for astrophysical observations, where the ranges
of scales are very often enormous.  At one extreme, magnetic fields of order 1 microgauss
at length-scales $\sim 1 Mpc$ in the intergalactic medium of galaxy clusters has been inferred 
from Faraday rotation measures \cite{Kronberg94,ZweibelHeiles97}. This technique, however,
gives only the magnetic field averaged over the line of sight. A bit closer to human scale, 
magnetic fields with kilogauss strength are observed at the surface of the sun by speckle 
interferometry and spectro-polarimetry with organization down to $\sim 200 km,$ the 
achievable spatial resolution of the techniques using current instruments 
\cite{Keller92,Stenflo13}. Indirect methods that combine modelling with observational 
data suggest, on the other hand, the ubiquitous presence of hidden, mixed-polarity fields on 
subresolution scales in the form of tangled, turbulent magnetic fields with an average strength 
of 100 gauss \cite{TrujilloBueno04,Stenflo13,Laggetal15}. Even the finest measurements of turbulent velocities in 
non-conducting fluids taken in terrestrial laboratories can presently resolve only to a Kolmogorov 
length-scale, e.g. $\mathcal{O}(0.1-1) mm$ in the recent experiments \cite{Zeffetal03,ThormannMeneveau14}, 
whereas there are theoretical arguments and numerical evidence for substantial velocity fluctuations at 
sub-Kolmogorov lengths \cite{Schumacher07,Yeungetal15}. 

The dynamics of high- and medium-density plasma flows is governed, nominally, by 
magnetohydrodynamic (MHD) equations \cite{Alfven50,Bellan06,Kulsrud05}. MHD is expected 
to apply well in the solar interior and photosphere,  for example, and approximately in the 
solar corona and in interstellar or intergalactic space. MHD also governs the motion of liquid metals
such as those used in industrial applications and those present in planetary cores. This model is given mathematically 
by a set of partial-differential equations (PDE's) for a continuous field of velocities, magnetic fields, 
densities, etc. Of course, the  fields that appear in this continuum description are an idealization 
of reality. All that can ever be observed experimentally are spatially-averaged or coarse-grained 
fields at some scale-resolution $\ell$. The fine-grained fields in conventional MHD are an 
appropriate idealization for plasma variables averaged over length-scales much greater than 
the electron and ion mean-free paths $\lambda_{e,i}$ and gyroradii $\rho_{e,i}$ but also much 
smaller than the gradient-lengths $\ell_\Delta\sim |\bu|/|\grad\bu|,|\bB|/|\grad\bB|$ set by the 
fluid-variable spatial variations.   However, the 
dynamics of coarse-grained field variables observed at resolution lengths $\ell\gg \ell_\Delta$ 
are {\it not} governed by conventional MHD equations, even if the fine-grained fields at 
length-scales $\ell_\Delta\gg\ell\gg \max\{\lambda_{e,i},\rho_{e,i}\}$ obey MHD  perfectly. 
The mode-elimination involved in coarse-graining leads to a ``renormalization''  of the 
conventional MHD equations that govern the fine-grained or ``bare'' field variables, and 
new terms appear in the coarse-grained dynamics that are due to nonlinear interactions
of the eliminated modes. 
 
This paper aims to give a systematic, theoretical approach to coarse-grained MHD.
There are several motivations to do so. First, it is impossible to give a consistent account 
of astrophysical observations without such a framework.  Paradoxes and quandries 
occur if one naively assumes that coarse-grained velocities and magnetic fields 
at large-scales are governed by conventional MHD. It is useful to have a dynamical 
framework that connects the directly observable variables. This is also very helpful from the 
point of view of numerical modelling, which is a tool of increasing importance in astrophysical applications 
(e.g. \cite{Gomezetal05,StoneGardiner07,Hawleyetal07,Jiangetal14,Shiokawaetal15,Teyssier15}). 
Because of the huge range of scales in nearly all astrophysical systems, however, numerical 
models cannot hope to resolve scales down to those where microscopic dissipation becomes effective.  
The coarse-graining
framework that we discuss here provides a basis for constructing models of the large-scales 
that faithfully reflect the MHD dynamics of the unresolved plasma fluid modes. In fact, the 
mathematical ``filtering formalism'' that we employ is the same as that  used in large-eddy 
simulation (LES) modelling of turbulent fluid flow, a technique currently under active 
development for MHD turbulence 
\cite{Agulloetal01,MuellerCarati02,HaugenBrandenberg06,
Chernyshovetal07,MieschToomre09, Chernyshovetal10,Mieschetal15}. 
The present work thus provides a theoretical counterpart to those computational efforts. 
This is also important since the fundamental basis of LES along with
the meaningfulness of scale-decompositions have been questioned \cite{Tsinober02}.
It is therefore useful to explain carefully the physical basis of the approach.

\subsection*{The cascade}
As an application of the coarse-graining approach, we derive necessary conditions required for the MHD non-linearities to  sustain a cascade of the three quadratic invariants, energy, magnetic helicity, and cross helicity.
At large Reynolds numbers, turbulent flows are characterized by the disordered and chaotic behavior of
fields in space and in time, a state which is called ``fully developed turbulence.'' It is inherently a phenomenon due to the non-linearities in the equations which couple motions at various
scales. One of its profound characteristics is the cascade of energy and other invariants across scales.

The prevailing understanding of the turbulent cascade owes its origin to Richardson (1926) \cite{Richardson26} who described
the energy transferred from eddies of size $\ell$ to eddies a fraction of that size, and so on. This cascade process transfers energy to successively smaller scales until it reaches scales at which viscous effects are able to efficiently dissipate kinetic energy into heat.

Therefore, the cascade  in turbulent flows acts as a ``bridge'' between the large inviscid scales and the small  viscous scales, such that it \emph{catalyzes}
the dissipation of energy residing in the large-scales. In hydrodynamics, it is a well established empirical fact \cite{Taylor35,Sreenivasan84,Pearsonetal02,Kanedaetal03} that 
the enhanced dissipation becomes independent of the value of viscosity at high Reynold's numbers. 
This is the so-called ``zeroth law of turbulence,'' which in hydrodynamic turbulence is described by 
the fundamental relation
\be \epsilon \sim \frac{u_{rms}^3}{L},
\lb{zeroth-law}\ee
first deduced by G. I. Taylor in 1935 \cite{Taylor35}. Here, $\epsilon$ is average energy dissipation (per unit mass) in a turbulent flow. It only depends on the {\it rms} velocity, $u_{rms}$, and the largest length-scale, $L$, in the flow.  

Building upon these ideas, Kolmogorov (1941) \cite{Kolmogorov41a} introduced the concept of an \emph{inertial range} of scales $\ell$,
which lie far from the largest and the smallest dissipative scales in the system: $L\gg \ell \gg \ell_\nu$. The flow at these scales evolves independently
of the particulars of the system such as the material making up the fluid, the microphysical properties such as viscosity, and the largest scales 
in the system such as geometry, boundary conditions, and the way the flow is stirred. Hence, Kolmogorov hypothesized a universal
behavior of the flow at such scales which solely evolve under their own internal dynamics.

The zeroth law implies that turbulent flows are capable of 
dissipating energy independently of viscosity, even if the latter is zero. 
This is known as the dissipative anomaly of Onsager \cite{Onsager49}, who pointed out that such a law requires that the velocity field be ``rough'' enough (with H\"older exponent 1/3 or less) to be able to sustain the non-linear transfer of energy to arbitrarily small scales. Onsager's result was first proved rigorously by Eyink \cite{Eyink94}, who relied on the equations of motion without using any closure (see also \cite{EyinkSreenivasan06,Eyink08}).

Formulating the problem in terms of inviscid dynamics is, of course, an idealization of the fact that  motions in the inertial range evolve and transfer energy to smaller scales without ``knowing'' anything about the existence of viscosity. Hence, fluid motion at such scales should be well-described by the inviscid Euler equation. In this paper, we show how such an approximation can be formalized rigorously within the coarse-graining framework. We will show how the approximation deteriorates at scales approaching the scale-range where viscosity has a measurable effect. The statement that the flow is able to dissipate energy in the limit of zero viscosity is equivalent to the statement  that the turbulent flow can sustain an energy cascade to \emph{arbitrarily small scales} until it is ultimately dissipated, in accordance with the zeroth law of turbulence.

Anomalous dissipation is not restricted to energy in Euler flows. Eyink \cite{Eyink2006b} showed that conservation of circulation (Kelvin's Theorem) can be violated in inviscid flows if the velocity field is rough enough. Similarly, magnetic reconnection, or the breakdown of magnetic flux conservation (Alfv\'en's Theorem), can occur in ``rough'' MHD flows in the absence of microphysical non-idealities as was shown in \cite{EyinkAluie06,Eyinketal13}. Those works showed that, in a sense, vorticity and magnetic flux, both being Lagrangian invariants, can also undergo a cascade. Recent numerical studies \cite{MininniPouquet09,DallasAlexakis14,Linkmannetal15} have also presented evidence in support of the existence of a zeroth-law for the total energy in MHD turbulence. All major theories of MHD turbulence \cite{Iroshnikov64,Kraichnan65,GoldreichSridhar95,Boldyrev05} implicitly rely on the existence of a dissipative anomaly.

In this paper, we will show that quadratic MHD invariants can be dissipated anomalously by ideal MHD if certain conditions on the roughness of the velocity and magnetic fields are met. We find that, while the conditions for the anomalous dissipation of energy and cross-helicity are easily satisfied in MHD turbulence, it is impossible for ideal MHD to dissipate magnetic helicity unless 3rd-order moments of either the velocity or magnetic field become infinite. In other words, energy and cross-helicity, but not magnetic helicity, can undergo a forward cascade to small dissipative scales in MHD at arbitrarily high Reynolds numbers.

In the spirit of this {\it Focus on} issue, we have tried to make the presentation 
accessible to a broad spectrum of readers by emphasizing the essential physics 
over technical details, while also attempting to maintain a certain level of mathematical rigor.
The contents of this paper are as follows. In section \ref{sec:FilteredDynamics}, we 
formulate the coarse-graining approach to incompressible MHD and discuss its connection with
other formalisms. 
In section \ref{sec:InertialDynamics}, we prove that microphysical mechanisms have negligible effect on the evolution of large scales.
In section \ref{sec:EnergyBudgets}, we derive budgets of kinetic and magnetic energy, both at scales larger and smaller than $\ell$, and discuss the conditions necessary for the energy cascade to operate at arbitrarily small scales. In section \ref{sec:MagneticHelicity}, we derive a large-scale budget of magnetic helicity, discuss the physical mechanism by which it is transferred between scales, and prove that its forward cascade to asymptotically small scales is impossible. In section \ref{sec:CrossHelicity}, we also derive a large-scale budget of cross helicity, discuss the physical mechanism by which it can cascade, and derive the necessary conditions for it to cascade to arbitrarily small scales. Section \ref{sec:Summary} summarizes the paper.

\section{Coarse-Grained Incompressible MHD\lb{sec:FilteredDynamics}}

\subsection{The Incompressible MHD Equations}

We consider plasmas that are well-described by the incompressible MHD equations
with velocity $\bu$ and magnetic field $\bB,$ in cgs units: 
\be \rho \partial_{t} \bu + \rho (\bu\cdot\grad)\bu = -\grad P + \frac{1}{c}\bJ\btimes\bB + \mu\nabla^{2}\bu
\lb{eq:u-eq} \ee
\be \partial_{t} \bB = \grad\times(\bu\times\bB) + \eta\nabla^{2}\bB
\lb{B-eq} \ee
\be \grad\cdot\bB=\grad\cdot\bu=0
\lb{eq:div-free} \ee
Here, mass density $\rho,$ shear dynamic viscosity $\mu$, and resistivity $\eta$, are all
assumed to be space-time constants. In what follows, we shall also use kinematic viscosity, $\nu=\mu/\rho$. The electric current $\bJ$ is given by the nonrelativistic 
approximation to Ampere's law as $\bJ= \frac{c}{4\pi}(\grad\btimes\bB)$. This allows the 
Lorentz force $\frac{1}{c}\bJ\btimes\bB$ to be rewritten as $ \frac{1}{4\pi}(\grad\btimes\bB)\btimes\bB 
= \frac{1}{4\pi}\grad\cdot(\bB\bB) -\grad(\frac{B^{2}}{8\pi}),$ so that the magnetic pressure 
$B^2/(8\pi)$ can be combined in eq.(\ref{eq:u-eq}) with the plasma pressure $P$ to give a total pressure 
$P_*=P+B^2/(8\pi).$ Eq.(\ref{B-eq}) for the magnetic field is a consequence of Faraday's
law, $\partial\bB/\partial t=-c\grad\btimes\bE$, and Ohm's law, 
\be \bE +\frac{1}{c}\bu\btimes\bB = \bJ/\sigma, \lb{Ohm} \ee
with the conductivity $\sigma=c^2/(4\pi\eta).$ 

We shall not discuss all assumptions underlying the validity of eqs.(\ref{eq:u-eq})-(\ref{eq:div-free}), 
but refer the reader to the literature for details (e.g.\cite{Bellan06}, Ch. 2 or \cite{Kulsrud05}, Ch. 3). 
We confine ourselves here to just a few relevant remarks. A fundamental assumption of the MHD 
approximation is that the mean-free-paths $\lambda_{j},\, j=i,e$ of ions and electrons 
(defined in terms of a Coulomb collision cross-section for a ``hard'' scattering event at 
an angle of $90^\circ$) must be much less than the ion and electron gyroradii $\rho_{j}=
m_jv_{\perp j}c/e_jB,\,j=i,e$. The latter, in turn, should be much less than a ``characteristic
length'' $\ell_\Delta$ associated with plasma flow variations. For the MHD approximation to be strictly 
accurate, this characteristic length should be taken to be a gradient-length of the plasma fluid variables  
$\ell_\Delta=\min\{|\bu|/|\grad\bu|,|\bB|/|\grad\bB|\}.$ Note that this length-scale could be much 
smaller than a large length-scale set by the macroscopic extent of the plasma. For example, 
in a turbulent MHD plasma very large gradients will form by nonlinear cascade, before 
they are damped by viscosity and resistivity. Here the fluid velocity $\bu$ is defined as the 
center-of-mass velocity of  a plasma element of dimension $\ell$ with $\max_{j=i,e}\{\lambda_j,\rho_j\}
\ll \ell \ll \ell_\Delta$ and, likewise, the magnetic field $\bB$ is averaged over a length-scale $\ell.$
Because of the assumed separation between the microscopic plasma length-scales 
$\lambda_j,\rho_j,\,j=i,e$ and the fluid-dynamical length-scale $\ell_\Delta$, the results of 
 averaging will not depend upon the particular length $\ell$ chosen in the allowed range.   

MHD eqs.(\ref{eq:u-eq})-(\ref{eq:div-free}), as written, involve additional approximations that are 
less essential and which were adopted largely for simplicity. For example, the simple version 
of Ohm's law adopted in eq.(\ref{Ohm}) could be made more realistic by the addition of a 
Hall term, electron pressure-gradient, anisotropic resistivity, etc. As we shall emphasize 
below, our conclusions do not depend on the particular simple form of Ohm's law in (\ref{Ohm}).
Another simplification that we have made is to assume flow incompressibility, which is appropriate 
for small Mach number. Although there would be some additional (interesting) complications, 
it should be possible to extend our analysis to compressible MHD \cite{Aluie13,Eyink2015}. The most realistic applications
of the present theory, where our specific assumptions are probably most well-satisfied, are 
to stellar interiors, the solar photosphere, and planetary dynamos.  

The incompressible MHD equations (\ref{eq:u-eq})-(\ref{eq:div-free}) have another formulation 
as a set of integral conservation laws:
\begin{eqnarray} 
&&\hspace{-2.5cm}
\rho\int_{\Omega} d^3 x\, \bu(\bx,t)-\rho\int_{\Omega} d^3 x\, \bu(\bx,t')
     =-\int_{t'}^t d\tau\, \int_{\partial \Omega}dA \,\,\hat{\bn}\bdot \left[P_*{\bf I} + 
         \rho\bu\bu-\frac{1}{4\pi}\bB\bB-\mu\grad\bu\right]_{\bx,\tau} \hspace{.5cm} \lb{u-eq-int}\\
&&\hspace{-2.5cm}
\int_{\Omega} d^3 x\, \bB(\bx,t)-\int_{\Omega} d^3 x\, \bB(\bx,t')
     = \int_{t'}^t d\tau\, \int_{\partial \Omega}dA \,\,\hat{\bn}\btimes \left[\bu\btimes\bB 
     -\eta(\grad\btimes\bB)\right]_{\bx,\tau} \lb{B-eq-int} \\
&&\hspace{-2.5cm}
 \int_{t'}^t d\tau\, \int_{\partial \Omega}dA \,\,\hat{\bn}\bdot \bB(\bx,\tau)= 
          \int_{t'}^t d\tau\, \int_{\partial \Omega}dA \,\,\hat{\bn}\bdot\bu(\bx,\tau) =0.
          \lb{div-free-int} 
\end{eqnarray} 
In eq. (\ref{u-eq-int}), ${\bf I}$ is the identity rank-2 tensor.

The integral formulation has a number of advantages, both physically and mathematically. 
Physically, (\ref{u-eq-int})-(\ref{div-free-int}) contain only observable variables and express
direct relations between them.  Eq.(\ref{u-eq-int}) equates the change of momentum in a region
$\Omega$ over the time-interval $[t',t]$ to the time-integrated momentum flux (stress) across the 
boundary $\partial \Omega$. Similarly, eq.(\ref{B-eq-int}) equates the change of magnetic field 
inside $\Omega$ over the interval $[t',t]$ to the time-integrated electromotive force (EMF) around 
the boundary surface. 
Finally, eq.(\ref{div-free-int}) states that there is zero net magnetic flux and mass flux across 
any closed surface. Assuming smoothness of solutions, the differential formulation, 
eqs.(\ref{eq:u-eq})-(\ref{eq:div-free}), is recovered as an idealization for small volume 
$vol(\Omega)$ and duration $\Delta t=t-t'.$ More precisely, if the diameter of $\Omega$ 
is much smaller than $\ell_\Delta=\max\{|\bu|/|\grad\bu|,|\bB|/|\grad\bB|\}$ and $\Delta t$ 
is much smaller than $\tau_\Delta=\max\{|\bu|/|\partial_t\bu|,|\bB|/|\partial_t\bB|\},$ then 
dividing eqs. (\ref{u-eq-int})-(\ref{div-free-int}) by the product $vol(\Omega)\cdot \Delta t$ 
gives back, effectively,  eqs. (\ref{eq:u-eq})-(\ref{eq:div-free}).

Mathematically, the integral formulation, eqs.(\ref{u-eq-int})-(\ref{div-free-int}), is more 
general than the differential formulation, eqs.(\ref{eq:u-eq})-(\ref{eq:div-free}). Although the two 
formulations are equivalent for {\it smooth} MHD solutions $\bu,\bB,$ the integral form makes 
sense even for {\it singular} (non-differentiable) solutions. Note that the integral 
eqs.(\ref{u-eq-int})-(\ref{div-free-int})
contain one fewer space/time derivative than the differential eqs.(\ref{eq:u-eq})-(\ref{eq:div-free}).
In fact, in the ideal limit $\nu,\eta\rightarrow 0,$  eqs.(\ref{u-eq-int})-(\ref{div-free-int}) contain
no derivatives at all and make sense for any square-integrable functions $\bu,\bB.$ 
The integral conservation laws (\ref{u-eq-int})-(\ref{div-free-int}) are equivalent to 
(\ref{eq:u-eq})-(\ref{eq:div-free}) interpreted in the {\it distribution (or weak) sense}, 
i.e. after smearing with a smooth, rapidly decaying test (or window) function, $\varphi(\bx,t):$  
\be \int d^3x\int dt \,\,\left[\rho (\partial_{t}\varphi) \bu 
 + (\grad\varphi)\bdot\left(P_*{\bf I}+\rho \bu\bu -\frac{1}{4\pi}\bB\bB\right)
 -\mu(\bigtriangleup\varphi)\bu\right]=\bzed
\lb{u-eq-dist} \ee
\be  \int d^3x\int dt \,\,\left[(\partial_{t}\varphi)\bB 
    +(\grad\varphi)\btimes(\bu\times\bB)
     -\eta(\bigtriangleup\varphi)\bB\right]=\bzed
\lb{B-eq-dist} \ee
\be \int d^3x\int dt \,\,(\grad\varphi)\cdot\bB=
      \int d^3x\int dt \,\,(\grad\varphi)\cdot\bu=0
\lb{div-free-dist} \ee
A Gaussian function, $\exp{\left(-|\bx|^2/2\right)}/(2\pi)^{3/2}$, is one possible choice of $\varphi(\bx,t)$. More detailed discussions of the integral formulation of PDEs can be found, for example, in \cite{Evans98}, Section 3.4 or in \cite{Leveque02}, Section 11.11.  By taking 
smooth test functions $\varphi(\bx,\tau)$ that more and more closely approximate the 
characteristic function $\chi_{\Omega\times [t,t']}(\bx,\tau)$, defined to be identically one over the spatio-temporal domain $\Omega\times [t,t']$ and zero otherwise, one can derive (\ref{u-eq-int})-(\ref{div-free-int}) from (\ref{u-eq-dist})-(\ref{div-free-dist}). 
Conversely, any smooth, rapidly decaying function can be approximated arbitrarily 
well by linear superpositions of characteristic (or top-hat) functions of smooth domains, $\varphi(\bx,\tau)
\approx \sum_i c_i \chi_{\Omega_i\times [t_i,t_i']}(\bx,\tau),$ and this allows 
(\ref{u-eq-dist})-(\ref{div-free-dist}) to be derived from (\ref{u-eq-int})-(\ref{div-free-int}) .  

Some of the ideas discussed here are quite old. It was emphasized by Bohr and 
Rosenfeld in the 1930's that only averaged magnetic fields $\int_\Omega d^3x\int_{t'}^t 
d\tau\,\bB(\bx,\tau)$ or smeared fields $\int d^3x\int d\tau \,\varphi(\bx,\tau)\bB(\bx,\tau)$
are physically meaningful. In the first place, it is only such averaged or smeared fields
which are experimentally measurable. See \cite{BohrRosenfeld33,BohrRosenfeld50}
for quantum electrodynamics and \cite{Rosenfeld51} for the case of classical electrodynamics
developed from the same point of view.  Furthermore, when the fields are singular, 
the smearing  is necessary to interpret the equations of motion distributionally. In the
quantum case, the singularity is implied by the canonical commutation relations of the 
local field operators, which contain Dirac delta functions. As is well-known, quantum 
fields are operator-valued distributions. However, singularities also appear in the 
classical fields in various physical limits, such as infinite-Reynolds number turbulence.
In that case, also, spatial-averaging or smearing with test functions is necessary.    

\subsection{The Coarse-Graining Approach}

It is useful to formalize the notion of a ``coarse-grained measurement'' of a field variable
$\ba(\bx)$ at a resolution length $\ell.$  Let us consider any kernel function $G$ which
is smooth, rapidly decaying, positive and with integral normalized to unity, $\int d^d r\, G(\br) 
= 1$, in $d$-dimensional space. Some of these requirements (e.g. smoothness, positivity) may be relaxed in 
some situations but there are, in any case, a wide class of functions that satisfy all four, such as the Gaussian function. 
Our results below hold for any choice of $G$ in this infinite class. We usually 
assume also that the kernel is centered, $\int d^d r \,\br \,G(\br) = \bzed,$ and that 
it has variance of order unity, $\int d^d r \,|\br|^2 G(\br) \approx 1.$
For any chosen length-scale $\ell$, we define a rescaled kernel $G_\ell(\br)=\ell^{-d} 
G(\br/\ell)$, so that normalization is preserved, 
$\int d^d r\, G_\ell(\br) = 1,$ but now $\int d^d r \,|\br|^2 G_\ell(\br) \approx \ell^2$, such
that most of the kernel's weight is in a ball of diameter $\approx\ell$, centered at $\bzed$.
The integral 
\be \fltr{\ba}_\ell(\bx) = \int d^d r\, G_\ell(\br) \ba(\bx+\br) \lb{filter} \ee
thus represents a local average of the ``fine-grained'' or ``bare'' field $\ba(\bx)$ over a spatial 
region of radius $\approx \ell$ centered around the point $\bx$ or a spatially ``coarse-grained'' 
field at length-scale $\ell.$ One can imagine that different kernels $G$ correspond to 
somewhat different types of experimental probes and measurement techniques. Similar 
mathematical methods may be applied to time-averaging or temporal coarse-graining, but, 
for reasons discussed more below, spatial coarse-graining suffices for our purposes 
without the need of any additional averaging over time, ensembles, etc. 

The coarse-graining operation (\ref{filter}), being a convolution, is linear and, therefore,
commutes with space (and time) 
derivatives. Coarse-grained incompressible MHD equations are easily derived from 
(\ref{eq:u-eq})-(\ref{eq:div-free}), as follows:
\be \rho\partial_{t} \ol{\bu}_\ell + \rho (\ol{\bu}_\ell\cdot\grad)\ol{\bu}_\ell = -\grad\ol{P}_{*\ell} 
+\frac{1}{4\pi} (\fB_{\ell}\bdot\grad)\fB_{\ell} -\grad\cdot\btau_\ell 
+\mu\nabla^{2}\ol{\bu}_\ell.
\lb{u-eq-ell} \ee
\be \partial_{t} \ol{\bB}_\ell = \grad\times(\ol{\bu}_\ell\times\ol{\bB}_\ell)  
+ c\grad\times\bepsilon_\ell+ \eta\nabla^{2}\ol{\bB}_\ell,
\lb{B-eq-ell} \ee
\be \grad\cdot\ol{\bB}_\ell=\grad\cdot\ol{\bu}_\ell=0.
\lb{div-free-ell} \ee
Eqs.(\ref{u-eq-ell})-(\ref{div-free-ell}) are the same as the original eqs. (\ref{eq:u-eq})-(\ref{eq:div-free}) 
but augmented by terms involving $\btau_\ell$ and $\bepsilon_\ell$ that represent the effect 
of the eliminated plasma fluid modes at scales $< \ell$. Eqs.(\ref{u-eq-ell})-(\ref{div-free-ell}) are exact and 
do not rely on asymptotic expansions. Furthermore, the coarse-grained equations are deterministic, describing the evolution of $\OL\bu_\ell(\bx;t)$ and $\OL\bB_\ell(\bx;t)$ at every location $\bx$ in the domain at every instant in time. Here  $\btau_\ell$ is the 
{\it subscale stress tensor} (or momentum flux) which describes the force exerted 
on scales larger than $\ell$ by fluctuations at scales smaller than $\ell$. 
It is the sum of two contributions, $\btau_\ell
=\btau_\ell^u+\btau_\ell^B$ where
\be \btau^{u}_\ell =\rho[ \ol{(\bu\bu)}_\ell -\fu_{\ell}\fu_{\ell}] \lb{R-stress} \ee
is the \emph {subscale Reynolds stress} and 
\be \btau^{B}_\ell = -\frac{1}{4\pi}[\ol{(\bB\bB)}_\ell - \fB_{\ell}\fB_{\ell}] \lb{M-stress} \ee 
is the {\it subscale Maxwell stress}. The former is the contribution to momentum transport
due to advection of subscale momentum by subscale velocity fluctuations, while the 
latter is the contribution that arises from the  contraction stress along lines of the 
subscale magnetic field. 
The other subscale term, in the coarse-grained induction equation,
\be \bepsilon_\ell = (1/c)[\ol{(\bu\btimes\bB)}_\ell -\ol{\bu}_\ell\btimes\ol{\bB}_\ell] 
\lb{EMF} \ee 
is the \emph{subscale electromotive force (EMF)}. This is (minus) the electric field generated by the 
motion of subscale magnetic field lines due to advection by subscale velocity fluctuations.  
In the coarse-grained Ohm's law, 
\be \ol{\bE}_\ell +\frac{1}{c}\fu_\ell\btimes\fB_\ell =
                  -\bepsilon_\ell+ \fJ_\ell/\sigma, 
\lb{eq:FltrOhmsLaw}\ee                  
$\bepsilon_\ell$ acts as an effective ``non-ideality'' in addition to the contribution from the plasma 
conductivity.  It should be noted that the concept of the turbulent Reynolds stress dates back to 
Reynolds \cite{Reynolds1895} and in the context of coarse-graining, 
has been recognized since the early works in LES modelling \cite{Deardorff70,Leonard75}.
Similarly, the turbulent Maxwell stress and electromotive force are common 
in the mean field MHD literature (e.g. \cite{Moffatt78,KrauseRaedler80,Biskamp03}) which we briefly discuss below.

We emphasize that the coarse-grained MHD eqs.(\ref{u-eq-ell})-(\ref{div-free-ell}), along with subscale terms 
(\ref{R-stress})-(\ref{EMF}), are exact and do not involve any approximations 
or physical modeling beyond the original MHD eqs. (\ref{eq:u-eq})-(\ref{eq:div-free}). 
As such, eqs. (\ref{u-eq-ell})-(\ref{div-free-ell}) alone are not closed without 
knowledge of expressions (\ref{R-stress})-(\ref{EMF}) representing scales $<\ell$. 
Otherwise, a model for the subscale terms is needed as we discuss in 
section \ref{sec:TheoreticalDiscussion}.
Note that  the coarse-grained eqs. (\ref{u-eq-ell})-(\ref{div-free-ell}) for $\fu_\ell,\fB_\ell$ 
hold even if the ``bare''  fields $\bu,\bB$ are singular MHD solutions in the sense of distributions. 
Taking the smooth test function in eqs. (\ref{u-eq-dist})-(\ref{div-free-dist}) to have the specific form
$\varphi(\bx,t)=G_\ell(\bx-\bx')\psi(t),$ one obtains (\ref{u-eq-ell})-(\ref{div-free-ell}), valid
distributionally in time. In fact, the coarse-grained equations  (\ref{u-eq-ell})-(\ref{div-free-ell})
for all $\ell>0$ are mathematically equivalent to eqs. (\ref{u-eq-dist})-(\ref{div-free-dist}) for all 
smooth test functions  $\varphi.$           

In what follows, ``large-scale'' and ``small-scale'' denote spatial scales larger and smaller than $\ell$, respectively. It is important to keep in mind that such delineation of scales is variable and depends on $\ell$, which itself can be large or small relative the system's size. In fact, $\ell$ can be taken to be smaller than  viscous scale $\ell_\nu$ and/or resistive scale $\ell_\eta$, which may be important, for instance, when analyzing MHD flows at high or low magnetic Prandtl number, $Pm=\nu/\eta$. 
While $\ell$ within the coarse-graining formalism and the coarse-grained budgets we derive is not restricted to be larger than the microphysical scales, $\ell_\nu$ and/or $\ell_\eta$, we emphasize that some of the key results below (namely Propositions 1, 2 and Theorems 1, 2, 3) and their relevance depend on $\ell$ being larger than one or both of those scales. This will made clear where necessary.
Moreover, the reason for choosing a single coarse-graining scale $\ell$ in both momentum and induction equations (\ref{u-eq-ell})-(\ref{B-eq-ell}) is to keep the formalism simple and tractable, while highlighting that our analysis can be generalized to an entire hierarchy of scales as was done in \cite{EyinkAluie09,AluieEyink09,AluieEyink10}.
In the rest of this paper, we shall drop subscript $\ell$ from fields such as $\OL\bu_\ell$ when there is no risk for ambiguity.

\subsection{Theoretical Discussion\lb{sec:TheoreticalDiscussion}}

\subsubsection*{Mean-field approaches}
There is an obvious similarity of the coarse-grained MHD eqs. (\ref{u-eq-ell})-(\ref{div-free-ell}) 
with those of  {\it mean-field magnetohydrodynamics} which are commonly employed in dynamo 
theory, see \cite{KrauseRaedler80} or \cite{Biskamp03}, Chapter 4. In fact, the equations of 
mean-field MHD contain terms just like $\btau_\ell$ and $\bepsilon_\ell$ in ours, with similar 
physical meaning. However, there are important differences. The averaged fields $\langle\bu\rangle,
\langle\bB\rangle$ in mean-field MHD must in general be taken to be statistical ensemble-averages. 
This means that the equations of mean-field MHD do not hold for individual realizations of the ensemble, 
unlike the coarse-grained eqs. (\ref{u-eq-ell})-(\ref{div-free-ell}). Mean-field MHD may be derived 
by space-averaging for individual realizations only if there is a large separation in scales between 
the mean field and the fluctuations \cite{Moffatt78,KrauseRaedler80,Biskamp03,Suretal08,CattaneoHughes09}. 
Consider the gradient-length 
of the average fields $\bar{\ell}_\Delta=\min\{|\langle{\bu}\rangle|/|\grad\langle{\bu}\rangle|,
|\langle{\bB}\rangle|/|\grad\langle{\bB}\rangle|\}$ and the correlation-length $\ell_{corr}$ of the random 
fluctuations $\bu',\bB'.$ The validity of mean-field MHD for space-averages requires that $\bar{\ell}_\Delta
\gg \ell_{corr}.$ In that case, space-ergodicity allows coarse-grained fields to be replaced by 
ensemble-averages, $\fu_\ell\approx \langle\bu\rangle,\fB_\ell\approx \langle\bB\rangle,$ for 
$\ell$ in the range $\bar{\ell}_\Delta \gg \ell\gg \ell_{corr}$ and eqs. (\ref{u-eq-ell})-(\ref{div-free-ell})
for such $\ell$ reduce to those of mean-field MHD. The coarse-grained equations are thus a generalization 
of mean-field MHD, requiring no assumption of scale-separation (which is unrealistic in most contexts). 
A pertinent advantage of coarse-graining over mean-field approaches
is the freedom in choosing the specific spatial scales $\ell$ to probe, which 
allows for investigating physical processes at arbitrary spatial scales.
Mean-field frameworks, on the other hand, can only decompose fields 
into mean and fluctuating components without control over 
spatial scales. Moreover, the coarse-graining method 
of probing the dynamics is deterministic, allowing the description of the 
evolution of scales at every location and at every instant in time, whereas
the mean-field description is inherently statistical.

\subsubsection*{Macroscopic electromagnetism}
The spatial coarse-graining that we have employed is familiar in a number of related applications.
It is a standard tool used to derive macroscopic electromagnetism of continuous media starting 
with a microscopic description of charged particles. For example, see \cite{Russakoff70, 
Robinson73} and the textbook presentation of \cite{Jackson75}, Section 6.7. In this application,
spatial-averaging of microscopic charge and current densities coupled with gradient expansions 
yield the macroscopic polarization ${\bf P}$ and magnetization ${\bf M}$ (as well as higher-order 
multipole contributions).  
In that case, the small-scales contribution comes from lengths of the order of interparticle 
distances, implying a large separation of scales between the size of the 
molecules and the variation of the macroscopic fields, which results in a rapid convergence of the 
expansions. This is why in electromagnetism, the values for polarization ${\bf P}$ and magnetization ${\bf M}$ will not depend on the coarse-graining length $\ell$ (or spatial resolution), as long as it averages over scales greater than the  molecular correlation-length $\xi_{corr}$ and smaller than the macroscopic gradient length $\bar{\ell}_\Delta$.

As in macroscopic electromagnetism, our approach seeks a description of the small-scale 
plasma fluid modes as an ``effective medium'' acting on the large-scale modes. In our case,
effective terms $\btau_\ell$ and $\bepsilon_\ell$ are analogous to ${\bf P}$ and ${\bf M}$ in electromagnetism.
However, in the case of MHD turbulence, there is no separation of scales. The dynamics, in general, takes place over an entire continuum of scales, starting from that of the system, $L$, down to the microscopic length-scales. 
As a result, effective terms $\btau_\ell$ and $\bepsilon_\ell$ can be expected to vary as a function of $\ell$, {\it i.e.} the subscale terms in the coarse-grained equations are, in general, resolution dependent. 
For example, as we elaborate in section \ref{sec:BasicRelations}, the Maxwell stress, $\btau^B_\ell$, is proportional to magnetic energy at scales $<\ell$ and, therefore, can be expected to become smaller, in some average sense, with finer resolution $\ell$. This is because magnetic fluctuations intensity (as measured by their spectrum, for instance) typically decays with smaller $\ell$. On the other hand, the coarse-grained current, $c/4\pi \,\grad\btimes\OL\bB_\ell$, being a gradient of a non-smooth turbulent field, will typically increase in magnitude as resolution $\ell$ is made finer.

\subsubsection*{Large Eddy Simulation}
In the fluids context, the same averaging technique has been widely employed in the Large-Eddy 
Simulation (LES) modeling of turbulence in non-conducting fluids. See \cite{MeneveauKatz00} and, for a 
more theoretical discussion of the ``filtering approach" in turbulence, \cite{Germano92}. Our equations  
(\ref{u-eq-ell})-(\ref{div-free-ell}) coincide with those that are employed in LES modeling of MHD 
turbulence \cite{Agulloetal01,MuellerCarati02,HaugenBrandenberg06,Chernyshovetal07,MieschToomre09, Chernyshovetal10,Mieschetal15}. 
However, our use of
these equations will be rather different. In LES modelling, plausible but uncontrolled closures are 
adopted for the subscale terms $\btau_\ell$ and $\bepsilon_\ell.$ We shall here instead develop 
several exact estimates and some general physical understanding of these terms, which should 
provide constraints and insight into the scale-coupling to develop more physics-based closures.  Another 
difference is that LES generally takes the scale-parameter $\ell$ to be a fixed length of the order 
of the ``integral length-scale'' $L$ in the turbulent flow. Because the scales $\ell\ll L$ are expected 
to be universal in turbulent fluids, robustly accurate closures should be possible for those scales 
which would allow them to be modeled and effectively eliminated from the numerical simulation. 
To gain the greatest computational savings, LES practitioners generally take $\ell\approx L$. 
Our interest here is rather to understand the effective, coarse-grained MHD 
modes (\ref{u-eq-ell})-(\ref{div-free-ell}) for the {\it whole range of length-scales} $\ell,$ including 
various limits of $\ell$ both large and small.  

\subsubsection*{Renormalization Group}
In fact, there is a rather close connection of our analysis with renormalization-group (RG)
methods in statistical physics and quantum field-theory, especially ``real-space RG''
methods as discussed in \cite{Wilson75}, Section VI, and \cite{BurkhardtvanLeeuwen82}. 
The local, space-averaged field in (\ref{filter}) is like a ``block-spin'' in an equilibrium spin 
system and the coarse-grained eqs.(\ref{u-eq-ell})-(\ref{div-free-ell}) are analogous to 
an ``effective Hamiltonian/action'' for the block-spins with running coupling constants 
that depend upon the scale-parameter $\ell.$ [As an aside, we note that the 
coarse-grained eqs.(\ref{u-eq-ell})-(\ref{div-free-ell}) with random initial conditions or 
random forcing can be derived by path-integral methods through elimination of small-scale 
modes exactly as in RG analyses: see \cite{Eyink96} for the non-conducting fluid case.]  Our major 
concern in this paper is to understand how the subscale-generated terms $\btau_\ell$ 
and $\bepsilon_\ell$ in the ``renormalized'' equations depend upon length-scale $\ell$, 
by means of exact, nonperturbative estimates. We shall not make use here of technical 
RG-apparatus, such as rescalings, scale-differential equations, beta-functions,  etc. 
but the goal of our analysis is very similar.  

\subsection{Basic Relations and Observations\lb{sec:BasicRelations}}

We shall now develop basic estimates for the various terms in the coarse-grained eqs.(\ref{u-eq-ell})-(\ref{div-free-ell}). The coarse-graining operation  decomposes the fields into $\bu = \fu_{\ell} +  \bu' _{\ell}$ and  $\bB = \fB_{\ell} +  \bB' _{\ell}$, where ``prime'' denotes the residual part of the 
field at scales smaller than $\ell$. Using the definition (\ref{filter}) of $\OL\bu_\ell$ and $\OL\bB_\ell$, 
it is straightforward to verify that
\be \bu'_{\ell}(\bx) =-\langle \delta\bu(\bx;\br)\rangle_{\ell}\equiv -\int dr^3 G_\ell(\br)  \delta\bu(\bx;\br)  \lb{eq:RelIncrementsSmallScales_u}\ee
\be \bB'_{\ell}(\bx) =-\langle \delta\bB(\bx;\br)\rangle_{\ell}\equiv -\int dr^3 G_\ell(\br)  \delta\bB(\bx;\br)  \lb{eq:RelIncrementsSmallScales_b}\ee 
where 
$$\delta f(\bx;\br) = f(\bx + \br) - f(\bx)$$ 
is an increment of a field over a separation $\br$ at point $\bx$, and 
$$\langle \dots\rangle_{\ell} \equiv \int dr^3 G_\ell(\br)  (\dots)$$ 
is a volume average over separations $|\br| < \ell$, weighted by kernel $G$. Relations (\ref{eq:RelIncrementsSmallScales_u})-(\ref{eq:RelIncrementsSmallScales_b}) are exact and involve no approximation. They allow us to make a direct connection between small-scale fields arising from our coarse-graining approach with increments, which are common in turbulence theory.

The sub-scale terms  $\btau_\ell$ and  $\bepsilon_\ell$ appear in the coarse-grained 
equations to account for the effects of turbulent fluctuations at scales smaller than $\ell$.
It should be possible, therefore, to explicitly express them in terms of these fluctuations.
The sub-scale terms can indeed be expressed as a function of the fluctuations thanks to an exact identity discovered by Constantin et al. (1994) \cite{Constantinetal94} and physically explicated by Eyink (1995) \cite{Eyink95b}:
\begin{eqnarray}
\hspace{-1cm}\OL\tau_\ell(\bu,\bu) \hspace{.1cm} \equiv \hspace{.4cm}\btau_\ell^u/\rho&=& \ol{\bu\bu_{\ell}} -\fu_{\ell}\fu_{\ell}=
 \langle\delta_{r}\bu\delta_{r}\bu\rangle_{\ell}-\langle \delta_{r}\bu\rangle_{\ell}\langle \delta_{r}\bu\rangle_{\ell}\lb{identity}\\ 
\hspace{-1cm}\OL\tau_\ell(\bB,\bB) \equiv-4\pi\,\btau_\ell^B&=&  \ol{\bB\bB_{\ell}} - \fB_{\ell}\fB_{\ell}=
\langle\delta_{r}\bB\delta_{r}\bB\rangle_{\ell}-\langle \delta_{r}\bB\rangle_{\ell}\langle \delta_{r}\bB\rangle_{\ell}\\
\hspace{1.6cm}c\,\bepsilon_\ell&=& \ol{(\bu\btimes\bB)}_\ell -\ol{\bu}_\ell\btimes\ol{\bB}_\ell=
\langle\delta_{r}\bu\btimes\delta_{r}\bB\rangle_{\ell}-\langle \delta_{r}\bu\rangle_{\ell}\btimes\langle \delta_{r}\bB\rangle_{\ell}\lb{EMF-identity}
\end{eqnarray}
In fact, for any two fields $f$ and $g$, their sub-scale nonlinear contribution is
\be \OL\tau_\ell(f,g) \equiv \OL{f\,g}_\ell - \OL{f}_\ell\,\OL{g}_\ell.
\ee
We shall reserve notation $\btau_\ell = \btau^u_\ell+\btau^b_\ell$ for the subscale Reynolds and Maxwell stresses.

Using $\int dr^3 \grad G(\br) = 0$ and integration by parts, spatial derivatives of coarse-grained fields can also be expressed in terms of increments as
\be \grad\ol{f}_{\ell} = -\frac{1}{\ell}\int dr^3 (\grad G)_\ell(\br) ~\delta f(\bx;\br). \ee
From the above considerations,  rigorous big-$O$ upper bounds can be derived for various terms in the coarse-grained  equations (\ref{u-eq-ell}),(\ref{B-eq-ell}):
\be \grad\bdot\btau_\ell =     O {\bigg(}\frac{\rho|\delta\bu(\ell)|^{2} + |\delta\bB(\ell)|^{2}/4\pi}{\ell} {\bigg)} \lb{eq:taubounds}\ee
\be \grad\times\bepsilon_\ell = O {\bigg(}\frac{|\delta\bu(\ell)| |\delta\bB(\ell)|}{\ell} {\bigg)} \lb{eq:epsilonbounds}\ee
\be \nu\nabla^{2}\ol{\bu}_\ell =     O {\bigg(}\frac{|\delta\bu(\ell)|^{2}}{\ell}Re^{-1}_{\ell} {\bigg)} \lb{viscousterm}\ee
\be  \eta\nabla^{2}\ol{\bB}_\ell = O {\bigg(}\frac{|\delta\bB(\ell)|^{2}}{\ell}Lu^{-1}_{\ell} {\bigg)} \lb{resistiveterm}\ee
where $Re_{\ell} \equiv \frac{\delta u(\ell) \ell}{\nu} $ and $Lu_{\ell} \equiv \frac{\delta B(\ell) \ell}{\eta}(4\pi\rho)^{-1/2}$ are the Reynolds and Lundquist numbers, respectively, at scale $\ell$. For rigorous details, see Eyink \cite{Eyink05}.
These bounds make it apparent that the nonlinear subscale terms dominate over microphysical terms in the dynamics of large scales $\ell$. We shall discuss this in more depth below.

\section{Inertial Range Inviscid Dynamics\lb{sec:InertialDynamics}}
Rewriting the MHD equations, filtered at a large scale $\ell\gg \max\{\ell_\nu,\ell_\eta\}$, 
with the viscous and resistive terms dropped, 
\be \rho\partial_{t} \ol{\bu}_\ell + \rho (\ol{\bu}_\ell\cdot\grad)\ol{\bu}_\ell = -\grad\ol{P}_{*\ell} 
+\frac{1}{4\pi} (\fB_{\ell}\bdot\grad)\fB_{\ell} -\grad\cdot\btau_\ell 
\lb{u-Ideal-ell} \ee
\be \partial_{t} \ol{\bB}_\ell = \grad\times(\ol{\bu}_\ell\times\ol{\bB}_\ell)  
+ c \grad\times\bepsilon_\ell,
\lb{B-Ideal-ell} \ee
it becomes clear that eqs. (\ref{u-Ideal-ell}),(\ref{B-Ideal-ell}) are identical to the 
\emph{ideal} MHD equations,
\be \rho\partial_{t} \bu + \rho (\bu\cdot\grad)\bu = -\grad{P}_{*} 
+\frac{1}{4\pi} (\bB\bdot\grad)\bB 
\lb{u-Ideal} \ee
\be \partial_{t} \bB = \grad\times(\bu\times\bB),
\lb{B-Ideal} \ee
filtered at the scale $\ell$. The following proposition, which was derived in collaboration with Gregory L. Eyink,
 shows this rigorously. To avoid additional complications due to boundaries, we consider 
 a domain $\mathbb{T}^3=[0,L_{\mbox{\tiny{dom}}})^3$ that is periodic.

\begin{Prop} If solution $(\bu,\bB)$ of the non-ideal MHD equations (\ref{eq:u-eq})-(\ref{eq:div-free}) over a periodic domain $\mathbb{T}^3$ have finite energy, then non-ideal terms in the large-scale
dynamics (\ref{u-eq-ell}),(\ref{B-eq-ell}) vanish at every point $\bx$ as $\nu,\eta\to0$.
\lb{Prop1}\end{Prop}
{\it Proof of Proposition \ref{Prop1}:}
Using integration by parts, viscous diffusion in the coarse-grained momentum equation (\ref{u-eq-ell}) can be written as
\begin{eqnarray}
\nu~ \nabla^2 \OL{\bu}_\ell(\bx) 
&=&\frac{\nu}{\ell^2}\int d\br ~(\nabla^2 G)_\ell(\br) ~\bu(\bx+\br), \nonumber
\end{eqnarray}
where $(\partial_i G)_\ell(\br) = \ell^{-3}\partial G(\br/\ell)/\partial (r_i/\ell)$. 
This can be bounded using the Schwartz inequality,
\begin{eqnarray}
\bigg|\nu ~\nabla^2 \OL{\bu}_\ell(\bx) \bigg | 
&\le& \frac{\nu}{\ell^2}\int d\br \bigg |(\nabla^2 G)_\ell(\br) ~\bu(\bx+\br)    \bigg | \nonumber\\
&\le& \frac{\nu}{\ell^2} ~V^{\frac{1}{2}} ~\big\|(\nabla^2 G)_\ell \big\|_2 ~V^{\frac{1}{2}}\big\|\bu \big\|_2  \nonumber\\
&=&\frac{\nu}{\ell^2}~u_{rms} ~\bigg(\frac{L_{\mbox{\tiny{dom}}}}{\ell}\bigg)^{\frac{3}{2}} 
\bigg(\int d{\bf s} \big|\nabla^2 G({\bf s})\big|^2 \bigg)^{\frac{1}{2}}
\nonumber\end{eqnarray}
where $\|\dots\|_p = \langle|\dots|^p\rangle^{1/p}$ is the $L_p$-norm, $\langle\dots\rangle=\frac{1}{V}\int d^d\bx(\dots)$ is a space average, $L_{\mbox{\tiny{dom}}}^3 = V$ is the domain's volume, and ${\bf s} = \br/\ell$ is a dimensionless vector. $u_{rms}$ is the root-mean-square, $\langle |\bu|^2\rangle^{1/2}$.
Since $G(\br)$ is smooth, its derivatives are uniformly bounded and we have $\big\|\nabla^2 G \big\|_2 = \const < \infty$. The same argument applies to $\eta \nabla^2 \fB_\ell$. We remark that for spatially compact filtering kernels, $\nu~ \nabla^2 \OL{\bu}_\ell(\bx)$  depends on the flow only within a region of size $\mathcal{O}(\ell)$ around $\bx$. Therefore, Proposition \ref{Prop1} can be easily extended to the infinite domain, $\mathbb{R}^3$, by considering a local region of size, $L_{\mbox{\tiny{reg}}}$, centered at $\bx$ that is sufficiently larger than scale $\ell$ to replace $L_{\mbox{\tiny{dom}}}$ in our bounds, and assuming \emph{locally} finite energy within this region.

Our bound implies that the viscous and resistive terms, $\nu ~\nabla^2 \OL{\bu}_\ell(\bx)$ and $\eta ~\nabla^2 \OL{\bB}_\ell(\bx)$, vanish at every point in space in the limit of vanishing viscosity and resistivity, respectively. 
The fact that non-ideal effects are negligible everywhere in the domain is a significantly stronger conclusion than what can be obtained with more traditional scale-analysis methods, such Fourier analysis, which treat the flow in a globally averaged sense. Our bound also implies that for a fixed viscosity and resistivity, the \emph{direct} contributions from microphysical processes on the dynamics of scales larger than $\ell$ have a vanishing role for larger\footnote{Here, one can take both $\ell\to\infty$ and $L_{\mbox{\tiny{dom}}}\to \infty$ while keeping their ratio constant, $L_{\mbox{\tiny{dom}}}/\ell=\const$. Alternatively, one can fix $L_{\mbox{\tiny{dom}}}$ and consider $\ell\to L_{\mbox{\tiny{dom}}}$.}   $\ell$. We emphasize the word ``direct,'' because such large scales can, in principle, be linked \emph{indirectly} to microphysical processes at small scales via the nonlinearities, $\btau$ and $\bepsilon$ in eqs. (\ref{u-eq-ell}),(\ref{B-eq-ell}), which can couple different scales. The extent to which disparate length scales are coupled to each other via the nonlinearities is referred to as scale (non-) locality and has been the subject of several recent studies \cite{DomaradzkiRogallo,YeungBrasseur,Zhou93a,Zhouetal96,Alexakisetal05a,Alexakisetal05b,Mininnietal05,Eyink05,Caratietal06,Mininni06,Domaradzki07a,Domaradzki07b,Mininni08,Domaradzkietal09,EyinkAluie09,AluieEyink09,AluieEyink10,Aluie11}.
Scale-locality depends inherently on the nonlinear physics, and the extent to which it holds in a flow can be directly probed using the coarse-graining framework \cite{Eyink05,EyinkAluie09,AluieEyink09,AluieEyink10,Aluie11}.

Proposition \ref{Prop1} expresses the important 
fact that scales in the inertial range $L\gg\ell\gg\max\{\ell_\nu,\ell_\eta\}$ are governed by ideal MHD dynamics supplemented with additional nonlinear terms accounting for the inter-scale coupling. The direct role of microphysical processes 
in the evolution of these scales is negligible at high Reynolds numbers. The coarse-graining approach formalizes such ideas in a precise and transparent manner.

\section{Energy Budgets\lb{sec:EnergyBudgets}}
\subsection{Large-Scale Kinetic Energy}
The kinetic energy density balance for the large-scales may easily be derived from the filtered momentum equation (\ref{u-eq-ell}) and reads:
\be \partial_t (\rho\frac{|\fu|^2}{2}) + \partial_j\left[ (\rho\frac{|\fu|^2}{2} +\OL{P_*} ) \ol{u}_j  + \tau_{ij}\ol{u}_i  -\frac{1}{4\pi}(\fu\cdot\fB)\ol{B}_j   - \nu\partial_j(\rho\frac{|\fu|^{2}}{2})\right] 
\lb{kinetic-large}\ee
$$=  -\Pi^u_\ell -\frac{1}{4\pi} \ol{B}_i\ol{B}_j\partial_j\ol{u}_i
 - \rho\nu|\grad\fu|^2.
$$
The divergence terms represent spatial transport of large-scale kinetic energy. Here, $ (\rho\frac{|\fu|^2}{2}) \fu$ describes the advection by the large-scale flow, $ \OL{P_*}\fu$ is transport resulting from large-scale pressure gradients,
$\btau\cdot\fu$ represents the \emph{turbulent diffusion} resulting from subscale fluctuations in the velocity and magnetic field, whereas $\nu\grad(\rho\frac{|\fu|^{2}}{2})$ is viscous diffusion due to random molecular motions.
In the presence of large-scale magnetic fields, energy can also be transported by large-scale traveling Alfv\'{e}n waves through $(\fu\cdot\fB)\ol{\bB}/4\pi$.

The first term on the right-hand side, $\Pi^u_\ell$, is usually called the sub-grid scale (SGS) dissipation or the SGS energy flux. It acts as a sink of large-scale kinetic energy, and is defined as
\be \Pi^u_\ell(\bx) \equiv -(\partial_{j}\ol{u}_{i})\tau_{ij} = -\rho(\partial_{j}\ol{u}_{i})\OL\tau_\ell(u_i,u_j) +(\partial_{j}\ol{u}_{i})\OL\tau_\ell(B_i,B_j)/4\pi . 
\lb{kinetic-flux}\ee
It has the physical meaning of ``deformation work'' due to the subscale Reynolds and Maxwell stresses acting against large scale straining motions of the flow \cite{TL}. Although $\Pi^u_\ell(\bx)$ is not sign-definite, numerical and observational evidence shows that it is positive in a space-average sense 
\cite{BiskampWelter89, Politanoetal89, Zhouetal04,Mininnietal06,MininniPouquet09}, and, therefore,
acts as a sink for the total energy at the large-scales. This kinetic energy is transferred to the small-scale plasma velocity, where $\Pi^u_\ell(\bx)$ acts as a source, as will be shown below, thus representing transfer of energy \emph{across} scales.

The second term on the RHS of eq. (\ref{kinetic-large}), $\frac{1}{4\pi}\fB^{T}\bdot\fS\bdot\fB$ when written in terms of 
the symmetric strain tensor, $\bS=(\grad\bu + \grad\bu^T)/2$,
 is the kinetic energy expended by the large-scale flow 
to bend and stretch the large-scale magnetic $\fB$-lines. It can be rewritten as 
$\frac{1}{4\pi}\fB^{T}\bdot\fS\bdot\fB = -\frac{1}{c} (\fJ\btimes\fB)\bdot\fu + \grad\bdot\left[\frac{1}{4\pi}\fB(\fu\bdot\fB) - \frac{|\fB|^2}{8\pi}\fu\right]$, with
the alternate interpretation of kinetic energy gained from acceleration of the plasma by the large-scale Lorentz force. In either form, this term appears with an opposite sign in the large-scale magnetic energy budget (\ref{magnetic-large})
and  represents an exchange of energy between the flow $\fu$ and the magnetic field $\fB$ at the large-scales. It does not transfer energy across scale $\ell$, which is why we describe it as \emph{large-scale conversion} between kinetic and magnetic forms of energy to be distinguished from SGS fluxes, $\Pi^u_\ell$ and $\Pi^b_\ell$, which transfer energy \emph{across} scales.

The last term, $ - \rho\nu|\grad\fu|^2$, on the RHS of eq. (\ref{kinetic-large}) is negative semi-definite, accounting for the direct dissipation of  large-scale kinetic energy density by molecular viscosity of the fluid. It is smaller than the kinetic energy flux $\Pi^u_\ell(\bx)$ by a factor of the Reynolds number at scale $\ell$, and therefore is negligible with $\ell$ is large or $\nu$ is small. The following proposition, which was derived in collaboration with Gregory L. Eyink, shows this rigorously.

\begin{Prop} 
If solution $(\bu,\bB)$ of the non-ideal MHD equations (\ref{eq:u-eq})-(\ref{eq:div-free}) 
over a periodic domain $\mathbb{T}^3$ have finite energy, 
then viscous and resistive dissipation in the large-scale energy budgets (\ref{kinetic-large}) and (\ref{magnetic-large}) vanish at every point $\bx$ as $\nu\to 0$ and $\eta\to 0$, respectively.
\lb{Prop2}\end{Prop}
{\it Proof of Proposition \ref{Prop2}:}
Using integration by parts, viscous dissipation in the large-scale kinetic energy budget (\ref{kinetic-large}) can be rewritten as
\begin{eqnarray}
\hspace{-2cm}\nu~ \grad \OL{\bu}_\ell(\bx) {\bf :} \grad \OL{\bu}_\ell(\bx) 
&=&\frac{\nu}{\ell^2}\int d\br_1 ~(\grad G)_\ell(\br_1) ~\bu(\bx+\br_1) ~{\bf :}\int d\br_2 ~(\grad G)_\ell(\br_2) ~\bu(\bx+\br_2), \nonumber
\end{eqnarray}
where $(\partial_i G)_\ell(\br) = \ell^{-3}\partial G(\br/\ell)/\partial (r_i/\ell)$. 
This can be bounded using the Schwartz inequality,
\begin{eqnarray}
\hspace{-2cm}\nu \,|\grad \OL{\bu}_\ell|^2(\bx) 
&\le& \frac{\nu}{\ell^2}\int d\br_1 \bigg |(\partial_j G)_\ell(\br_1) ~u_i(\bx+\br_1) \bigg | \int d\br_2 \bigg |(\partial_j G)_\ell(\br_2) ~u_i(\bx+\br_2)    \bigg | \nonumber\\
&\le& \frac{\nu}{\ell^2} ~V ~\big\|(\grad G)_\ell \big\|^2_2 ~V~\big\|\bu \big\|^2_2  \nonumber\\
&=&\frac{\nu}{\ell^2} ~\bigg(\frac{L_{\mbox{\tiny{dom}}}}{\ell}\bigg)^{3}\big\|\bu\big\|^2_2 
~\int d{\bf s} \big|\grad G({\bf s}) \big|^2 
\nonumber\end{eqnarray}
Since $G(\br)$ is smooth, its derivatives are uniformly bounded and we have $\big\|\grad G \big\|_2 = \const < \infty$. The same argument applies to $\eta |\grad \OL{\bB}_\ell|^2$ in the large-scale magnetic energy budget (\ref{magnetic-large}) below. Therefore, for a fixed $\ell$, direct dissipation of energy at scales larger than $\ell$ vanishes everywhere in the domain in the limit of small viscosity and resistivity. Equivalently, for fixed values of $\nu$ and $\eta$, direct viscous/resistive dissipation acting at scales larger than $\ell$ vanishes everywhere in the domain for large $\ell$. Similar to Proposition \ref{Prop1}, Proposition \ref{Prop2} can also be extended to the infinite domain, $\mathbb{R}^3$.
 
\subsection{Small-Scale Kinetic Energy}
A corresponding small-scale energy or ``subscale kinetic energy''  may be defined as
\be
\ol{k}_\ell\equiv \frac{\rho}{2}\ol{\tau}_\ell(u_i,u_i). 
\lb{band-k} \ee
It is a positive quantity at every point in the flow if and only if the filtering kernel $G(\br)$ is positive 
for all $\br$, as was proved by Vreman et al. 1994 \cite{Vremanetal94}. Indeed, it can be rewritten 
as $\int d\br~G(\br) \frac{1}{2}|\bu(\bx+\br) - \ol{\bu}(\bx)|^2,$ which is the energy density averaged 
over a region of size $\ell$ around $\bx$ in a frame co-moving with the local large-scale velocity 
$\ol{\bu}(\bx)$  \cite{Vremanetal94}. Furthermore, integrating $\frac{1}{2}\ol{\tau}(u_i,u_i) $ in space 
gives $\int d\bx~ \frac{1}{2} |\bu(\bx)|^2- \int d\bx~ \frac{1}{2} |\ol{\bu}(\bx)|^2$, which is the total energy 
less the energy at large scales. It is straightforward to derive the energy budget of the small scales as:
\begin{eqnarray} 
\lefteqn{ \partial_t \frac{\rho}{2}\ol{\tau}_\ell(u_i,u_i)  + \partial_{j} \bigg[ \frac{\rho}{2}\ol{\tau}_\ell(u_i,u_i)\ol{u}_j + \ol{\tau}_\ell(P+\frac{|B|^2}{8\pi},u_j) + \frac{\rho}{2}\ol{\tau}_\ell(u_i,u_i,u_j)  } \nonumber\\ 
& &{}- \frac{1}{4\pi}\ol{\tau}_\ell(u_i,B_i,B_j) -\frac{1}{4\pi}\ol{\tau}_\ell(u_i,B_j)\ol{B}_i -\frac{1}{4\pi}\ol{\tau}_\ell(u_i,B_i)\ol{B}_j -\rho\nu\partial_j \frac{1}{2}\ol{\tau}_\ell(u_i,u_i) \bigg] \nonumber\\ 
& &{}= \Pi^u_\ell  -\frac{1}{4\pi}( \ol{B_i B_j\partial_j u_i}- \ol{B}_i\ol{B}_j\partial_j\ol{u}_i)  - \nu\ol{\tau}_\ell(\partial_iu_j,\partial_iu_j).
\lb{eq:kinetic-small}\end{eqnarray}
A similar budget in pure hydrodynamics was derived by Germano \cite{Germano92}. Here, 
$\ol{\tau}_\ell(f,g,h) \equiv \OL{fgh}_\ell -\OL{f}_\ell \OL{\tau}_\ell(g,h)
-\OL{g}_\ell \OL{\tau}_\ell(f,h)-\OL{h}_\ell \OL{\tau}_\ell(f,g)-\OL{f}_\ell\OL{g}_\ell\OL{h}_\ell$, for 
fields $f$, $g$, and $h$, which we interpret as the advection of subscale kinetic energy due to turbulent
fluctuations at scales $<\ell$. The first term inside the divergence  represents the advection of small-scale kinetic energy by the large scale flow, while the second term involves transport due to subscale pressure. The third and fourth terms are due to turbulent diffusion. The fifth and sixth terms inside the divergence are transport due to the large-scale magnetic field, and the last term is viscous diffusion.

The energy flux $\Pi^u_\ell(\bx)$ on the RHS now acts as source for the small-scale flow, representing the energy transferred from scales larger than $\ell$ at point $\bx$. The following two terms in parentheses, on the RHS of 
eq. (\ref{eq:kinetic-small}), represent kinetic-magnetic energy \emph{conversion} at small-scales, and do not contribute to the transfer of energy \emph{across} scales. The same two terms appear with an opposite sign in the small-scale magnetic energy budget (\ref{magnetic-small}).
The last term on the RHS is direct viscous dissipation of small-scale kinetic energy, which, unlike the viscous term in the large-scale budget (\ref{kinetic-large}), does not have to vanish in the limit of small viscosity. Indeed, it follows from the zeroth law of turbulence that it is of a magnitude equal to $\Pi^u_\ell(\bx) -( \ol{B_i B_j\partial_j u_i}- \ol{B}_i\ol{B}_j\partial_j\ol{u}_i)/4\pi$, on average.

\subsection{Large-Scale Magnetic Energy}
The rest of the system's energy is in the magnetic field, whose large-scale budget is:
\begin{eqnarray} 
\lefteqn{ \partial_t(\frac{|\fB|^2}{8\pi}) + \partial_j \left[(\frac{|\fB|^2}{8\pi}) \ol{u}_j  - \frac{c}{4\pi}(\bepsilon\btimes\fB)_j - \eta\grad(\frac{|\fB|^{2}}{8\pi})\right]   } \nonumber\\ 
& &{}= -\Pi^b_\ell+ \frac{1}{4\pi}  \ol{B}_i\ol{B}_j\partial_j\ol{u}_i - \frac{1}{4\pi} \eta|\grad\fB|^2.
\lb{magnetic-large}\end{eqnarray}
The first term inside the divergence in eq. (\ref{magnetic-large}) is magnetic energy advected by the large-scale flow, the second term is the \emph{turbulent Poynting flux} which transports magnetic energy with the aid of the turbulent electric field $-\bepsilon$, while the last term is diffusive transport due to microphysical resistivity.

The first term on the RHS is the magnetic SGS flux, defined as
\be \Pi^b_\ell \equiv - \fJ\cdot\bepsilon = -\frac{1}{4\pi} \partial_j\ol{B}_i [ \tau(u_j,B_i) - \tau(u_i,B_j) ].
\lb{magnetic-flux}\ee
It is a \emph{turbulent Ohmic dissipation} accounting for large-scale magnetic energy that is lost to scales smaller than $\ell$.
Note that $-\bepsilon$ is an electric field generated by the turbulence as it enters equation (\ref{EMF}) of the coarse-grained Ohm's law. 
Similar to the ``deformation work'' in (\ref{kinetic-large}), this term, while not sign-definite, is positive on average, 
as a result of the dynamics \cite{BiskampWelter89, Politanoetal89, Zhouetal04,Mininnietal06,MininniPouquet09}. 
Geometrically, this implies that in a turbulent plasma, the large-scale current $\fJ$ tends to anti-align with the 
turbulent EMF, $\bepsilon$, on average, at all inertial scales $\ell$.

The second term on the RHS of eq. (\ref{magnetic-large}) is the energy gained by the large-scale magnetic field  as it is bent and stretched by the large-scale plasma flow. It is a \emph{conversion term} that balances exactly a corresponding term in the large-scale kinetic budget eq. (\ref{kinetic-large}), and does not involve energy transfer across scale $\ell$.
The last term on the RHS of (\ref{magnetic-large}) is direct dissipation of  large-scale magnetic energy
 due to Spitzer resistivity.
It is smaller than the magnetic flux $\Pi^b_\ell$ by a factor of the magnetic Reynolds number at scale $\ell$. 
Therefore, as shown rigorously in Proposition \ref{Prop2}, it is negligible everywhere in the domain when $\ell$ is large or $\eta$ is small.

\subsection{Small-Scale Magnetic Energy}
The corresponding  magnetic energy budget for scales smaller than $\ell$ reads
\begin{eqnarray} 
 \partial_t \frac{1}{8\pi}\ol{\tau}_\ell(B_i,B_i)  
 & & + \partial_{j} \bigg[ \frac{1}{8\pi}\ol{\tau}_\ell(B_i,B_i)\ol{u}_j +\frac{1}{4\pi}\ol{\tau}_\ell(u_i,B_j)\ol{B}_i \nonumber\\
 && \hspace{.6cm}+ \frac{1}{8\pi}\ol{\tau}_\ell(B_i,B_i,u_j) -\eta\partial_j \frac{1}{8\pi}\ol{\tau}_\ell(B_i,B_i) \bigg]  \nonumber\\ 
&  & = \Pi^b_\ell  +\frac{1}{4\pi}( \ol{B_i B_j\partial_j u_i}- \ol{B}_i\ol{B}_j\partial_j\ol{u}_i)  - \frac{\eta}{4\pi}\ol{\tau}_\ell(\partial_iB_j,\partial_iB_j).
\lb{magnetic-small}\end{eqnarray}
The magnetic energy flux $\Pi^b_\ell(\bx)$ acts as source for the small-scale field, when it acts as a sink for the large-scale $\fB$
in (\ref{magnetic-large}), and accounts for the magnetic energy transferred \emph{across} scale $\ell$ at location $\bx$ in the domain.
The second two terms on the RHS of eq. (\ref{magnetic-small}) exactly balance with corresponding terms in the small-scale kinetic budget (\ref{eq:kinetic-small}),
and represent the kinetic-magnetic energy \emph{conversion}  at the small-scales, 
without involving energy transfer across scale $\ell$. 
The last term on the RHS is direct resistive dissipation, which balances, on average, 
the net energy provided to the small-scales by the flux $\Pi^b_\ell$ and conversion
$\frac{1}{4\pi}( \ol{B_i B_j\partial_j u_i}- \ol{B}_i\ol{B}_j\partial_j\ol{u}_i)$.

The first term inside the divergence  represents the advection of small-scale magnetic energy by the large scale flow, while the second term involves transport with the aid of the large-scale field $\fB$. The third term accounts for turbulent diffusion and the last term is resistive diffusion.

\subsection{Turbulent Energy Dissipation}
After dropping the viscous and resistive dissipation terms from eqs. (\ref{kinetic-large}),(\ref{magnetic-large}), which we have proved to be negligible for sufficiently large $\ell$, or small $\nu$ and $\eta$, the total energy budget reads,
\begin{eqnarray} 
\lefteqn{ \partial_t(\rho\frac{|\fu|^2}{2}+\frac{|\fB|^2}{8\pi}) } \nonumber\\ 
& &{}+ \partial_j \left[\big(\rho\frac{|\fu|^2}{2}+\frac{|\fB|^2}{8\pi}  +(\OL{P} + \ol{\frac{|\bB|^2}{8\pi}} )\big) \ol{u}_j   + \tau_{ij}\ol{u}_i  -\frac{1}{4\pi}(\fu\cdot\fB)\ol{B}_j - \frac{c}{4\pi}(\bepsilon\btimes\fB)_j \right]  \nonumber\\ 
& &{}= -D^{E}_{\ell},
\lb{Energy-ideal-ell}\end{eqnarray}
where $D^{E}_{\ell} =\Pi^u_\ell+\Pi^b_\ell$ is the total energy flux from scales $>\ell$ to scales $<\ell$.

An important property of the ideal MHD equations is that they do not have to conserve energy, in analogy to Onsager's dissipative anomaly
for Euler flows in pure hydrodynamics. This is because $D^{E} = \lim_{\ell\to0} (\Pi^u_\ell+\Pi^b_\ell)$ in the budget can be nonzero, expressing the fact that inertial range scales are able to transfer energy to \emph{arbitrarily small scales}, until it eventually
gets destroyed by microscopic processes. In order for the nonlinearities to cascade energy, the velocity and magnetic fields are required to be rough enough, as was proved by Caflisch et al.\cite{Caflisch97}.

Before stating the theorem, we shall present the essential physics behind Onsager's dissipative anomaly in hydrodynamic turbulence. First note that mean dissipation, $\nu\langle|\grad\bu|^2\rangle$ does not need to vanish as $\nu\to 0$ since velocity gradients increase without bound with increasing Reynolds numbers. This is because turbulent velocity is not a smooth field and gradients derive most of their contribution from the smallest fluctuating scales, of the same order as the viscous scale, $\ell_\nu$. As a result, a decrease in viscosity $\nu\to 0$ and, therefore, $\ell_\nu \to 0$, is offset by an increase in velocity gradients such that $\nu\langle|\grad\bu|^2\rangle =\epsilon$ remains constant, independent of viscosity or Reynolds number. This is a restatement of the zeroth law of turbulence, eq. (\ref{zeroth-law}).

While in any natural system, energy is ultimately dissipated by the microphysics, we have shown that viscosity is incapable of directly acting on the large scales. Hence the essential role of the nonlinear cascade in turbulence. This is represented by the flux term, which in pure hydrodynamics is $\Pi_\ell=\grad\OL\bu_\ell\,{\bf :}\,\btau^u_\ell$.
We have shown in section \ref{sec:BasicRelations} that each of $\grad\OL\bu_\ell$ and $\btau^u_\ell$ can be expressed in terms of increments. The upper bounds in section \ref{sec:BasicRelations} (see also \cite{Eyink05}) suggest that, 
\be
\Pi_\ell = \grad\OL\bu_\ell{\bf :} \btau^u_\ell \sim \mathcal{O}\left(\frac{\delta u(\ell)}{\ell} \cdot \delta u^2(\ell)\right),
\ee
where the symbol $\sim{\mathcal O}$ stands for ``same order of magnitude as.''
If velocity increments scale as 
\be \delta u(\ell) \sim \ell^{\sigma^u},
\lb{eq:HolderExponent}\ee
then 
\be
\Pi_\ell \sim \ell^{3\sigma^u -1}.
\lb{eq:PiHydroScalingHeuristic}\ee
The scaling exponent $\sigma^u$ is a measure of smoothness of the velocity field. To elucidate this fact, consider the Taylor series expansion, 
$$ \delta u (\ell) = u(x+\ell) - u(x) = \sum_n  \ell^n \frac{1}{n !}\nabla^n u
$$
as a function of $\ell$. As $\ell \to 0$, if the response of $\delta u (\ell)$ is to decrease in direct proportion to $\ell$, then it is an indication that the velocity at $x$ is smooth enough (differentiable) \emph{at that scale $\ell$}. If, on the other hand, $\delta u(\ell)$ is insensitive to $\ell$, {\it i.e} $\sigma^u=0$ in eq. (\ref{eq:HolderExponent}), then $u$ has a discontinuity at point $x$. Note that such a discontinuity is only at scale $\ell$ and does not have to be a ``true'' discontinuity in the abstract mathematical sense. Flow velocities in natural systems have a finite viscosity to smooth out such discontinuities at the smallest scales $\ell_\nu$. However, as Figure \ref{fig:discontinuity} illustrates, even if the velocity is infinitely differentiable in a natural setting, such that $\nabla^n u(x) < \infty$ for all $n$ in the Taylor expansion, when evaluating $\delta u (\ell)$ at large separations $\ell\gg\ell_\nu$, values for $\frac{1}{n !}\nabla^n u(x)$ may be so large that $\delta u (\ell)$ is insensitive to a decreasing $\ell$ over a wide range of scales $\ell$. Convergence of the Taylor series would only take hold when $\ell \lesssim \ell_\nu$. In real analysis, if $\sigma=1$, the function is said to be Lipschitz continuous and is generally characterized by corners. If a function's smoothness is intermediate between discontinuous and 
Lipschitz functions, it is said to be H\"older continuous with a H\"older exponent $0<\sigma<1$, and is generally characterized by cusps. The larger is the value of $\sigma$, the smoother is the field.

\begin{figure}
\centering
\includegraphics[totalheight=.25\textheight,width=.4\textwidth]{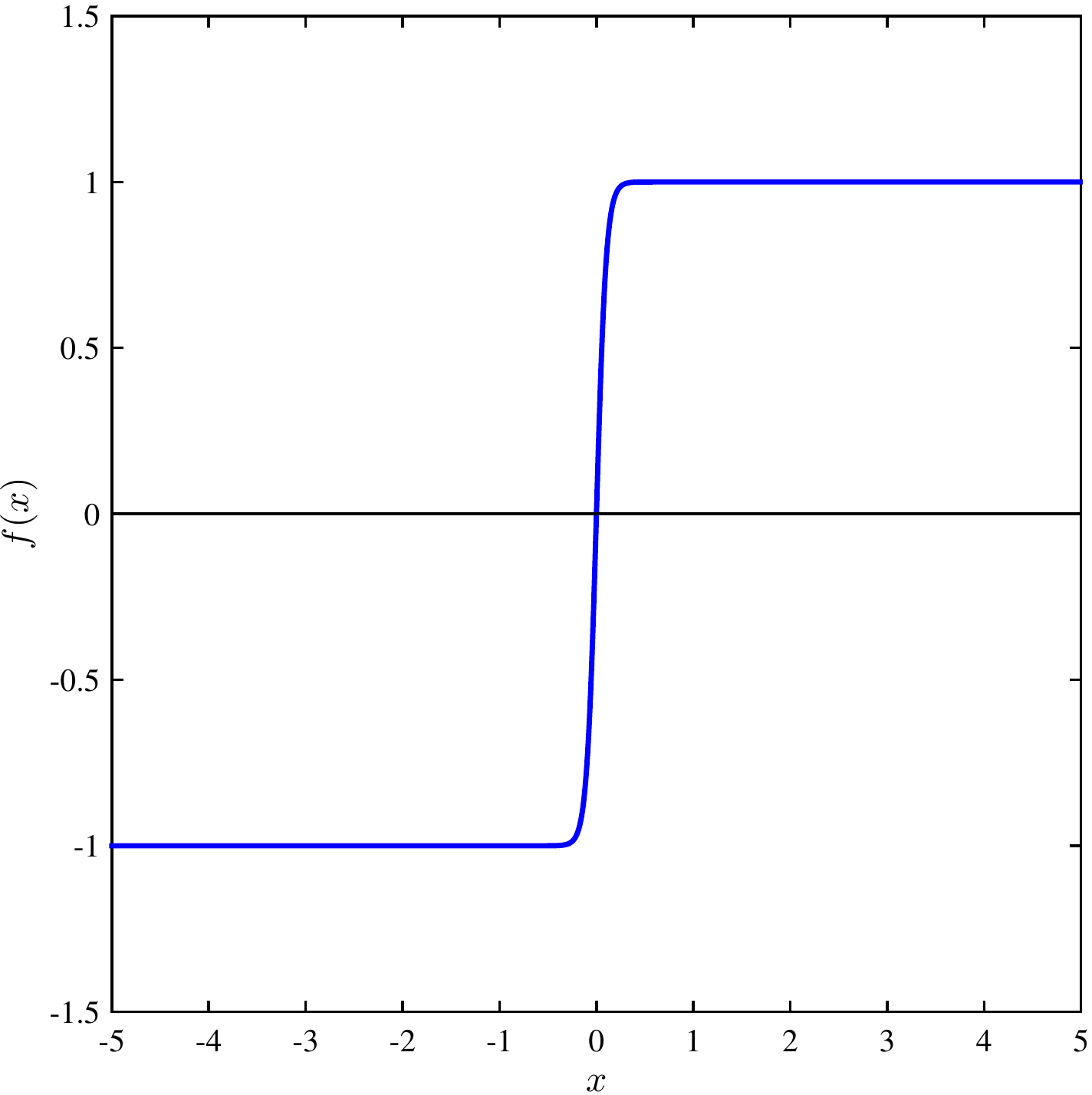}
\caption{The function $f(x) = -\tanh(-\frac{x}{\ell_\nu})$ with $\ell_\nu=0.1$ is smooth, but only at scales $\ell\lesssim \ell_\nu$. For larger separations, $\delta f(\ell; x=0)$ is insensitive to changes in $\ell$ and behaves as a discontinuous function. Therefore, $\delta f(\ell)\sim\ell^0$ at point $x$, for $\ell\gg\ell_\nu$.} 
 \lb{fig:discontinuity}
\end{figure}

Going back to eq. (\ref{eq:PiHydroScalingHeuristic}), if the velocity is smooth enough such that increments scale with $\sigma^u>1/3$, then $\Pi_\ell$ will shut down as $\ell\to 0$. At relatively moderate Reynolds numbers, such as in numerical simulations, $\Pi_\ell$ could still transfer a measurable amount of energy to the dissipation scales. However, if $\sigma^u>1/3$, the turbulent cascade would not persist to arbitrarily small scales as the Reynolds number increases. It is, therefore, imperative when analyzing simulation data, to scrutinize the trend as a function of Reynolds number and check if the phenomenon under study persists and can be extrapolated to the large Reynolds numbers present in nature. In actual incompressible hydrodynamic turbulence, $\sigma^u=1/3$, as implied from Kolmogorov's $4/5$-the law. This is the minimum roughness required to allow $\Pi_\ell$ to cascade energy to arbitrarily small scales.

Similar reasoning carries over to MHD turbulence, 
 $\Pi^u_\ell$ and  $\Pi^b_\ell$ can be expressed in terms of velocity and magnetic field increments. 
The velocity and magnetic structure functions are:
\begin{eqnarray}
\| \delta\bu(\br) \|_p &\sim& u_{rms} A_p (r/L)^{\sigma^u_p},\lb{scaling1}\\
\| \delta\bB(\br) \|_p &\sim& B_{rms} C_p (r/L)^{\sigma^b_p},\lb{scaling2}\\
\end{eqnarray}
for some dimensionless constants $A_p$ and $C_p$. 
Here, an $L_p$-norm $\| \cdot\|_p= \langle|\cdot|^p\rangle^{1/p}$ is the traditional
structure function $S_p=\langle|\cdot|^p\rangle$ raised to the $1/p$-th power.
For $p=2$, the exponents are related to the velocity and magnetic spectra, $E^u(k)\sim k^{-2\sigma^u_2-1}$ and $E^b(k)\sim k^{-2\sigma^b_2-1}$, respectively. Therefore, the steepness of a spectrum's slope indicates the smoothness of a field in a {\it root-mean-squared} sense. More generally, the scaling of structure functions in relations (\ref{scaling1})-(\ref{scaling2}) reflects the smoothness of fields in an $L_p$-normed sense. 

To derive the minimum roughness required to cascade energy in MHD, we first express the kinetic and magnetic energy fluxes in terms of increments:
\begin{eqnarray}
\Pi^u_\ell = -\grad\OL\bu_\ell{\bf :} (\btau^u_\ell+\btau^B_\ell) = O\left(\frac{\delta u(\ell)}{\ell} \cdot \left(\rho\delta u^2(\ell)+\delta B^2(\ell)/4\pi\right)\right),\\
\Pi^b_\ell = -\bJ\bdot\bepsilon_\ell = O\left(\frac{\delta B(\ell)}{\ell}\cdot \delta u(\ell)\,\delta B(\ell)\right).
\end{eqnarray}
If velocity and magnetic field increments scale as 
\be \delta u(\ell) \sim \ell^{\sigma^u}\mbox{\hspace{.5cm} and\hspace{.5cm}} \delta B(\ell) \sim \ell^{\sigma^b},
\ee
then 
\be
\Pi^u_\ell \sim \mathcal{O}\left(\min\left\{\ell^{3\sigma^u-1},\ell^{2\sigma^b+\sigma^u-1}\right\}\right) \mbox{\hspace{.5cm} and\hspace{.5cm}} \Pi^b_\ell \sim \mathcal{O}\left(\ell^{2\sigma^b+\sigma^u-1 }\right).
\ee
Therefore, if $\sigma^u>1/3$ and $2\sigma^b+\sigma^u>1$, both $\Pi^u_\ell$ and $\Pi^b_\ell$ will decay as $\ell\to 0$. In other words, if the nonlinearities in MHD are to cascade energy to arbitrarily small scales, either $\sigma^u\le1/3$ or $2\sigma^b+\sigma^u\le1$ (the 'or' is non-exclusive).
The following theorem by Caflisch et al.\cite{Caflisch97} restates this in mathematically rigorous manner\footnote{The original result in \cite{Caflisch97} was formulated over a periodic domain but the proof carries over to $\mathbb{R}^3$.}.

\begin{Th} (Caflisch et al. \cite{Caflisch97}) Let $\bu$ and $\bB$ be a weak solution of the ideal MHD equations over domain $\mathbb{T}^3$ or $\mathbb{R}^3$. Assume $\| \delta_{r}\bu\|_{3}  \sim r^{\sigma_{3}^{u}}$  and $\| \delta_{r}\bB\|_{3}  \sim r^{\sigma_{3}^{b}}$.
\begin{eqnarray}
\lefteqn{
\mbox{If\hspace{.5cm}} \sigma_{3}^{u} > \frac{1}{3} \mbox{\hspace{.5cm}and\hspace{.5cm}} \sigma_{3}^{u}+ 2\sigma_{3}^{b} > 1,
} \nonumber \\ 
& & \mbox{then\hspace{.5cm}}  \lim_{\ell\to 0} ~\left| \left\langle D^{E}_{\ell} \right\rangle \right | 
\le \lim_{\ell\to 0}  \left( \const~ \ell^{3\sigma_{3}^{u}-1} + \const~ \ell^{2\sigma_{3}^{b}+\sigma_{3}^{u}-1}  \right)
= 0, 
\nonumber \end{eqnarray}
where,
\begin{eqnarray}
D^{E}_{\ell} = \Pi^u_\ell + \Pi^b_\ell = -\grad\OL\bu_\ell{\bf :}\btau_\ell-\OL\bJ_\ell\bdot\bepsilon_\ell~.
\nonumber \end{eqnarray}
\lb{Th-Energy}\end{Th}

\section{Magnetic Helicity\lb{sec:MagneticHelicity}}

Magnetic helicity,  $H^{M}$,  is another quadratic invariant which can be transfered between scales. It was first discovered by Els\"asser in 1956 \cite{Elsasser56} and later, independently, by Woltjer in 1958 \cite {Woltjer58a}.
The topological significance of magnetic helicity
was realized by Moffatt (1969) \cite{Moffatt69}, who showed that it quantifies the degree of knottedness of  magnetic field lines in a system, measuring the number of links and twists 
between magnetic field loops. This topological interpretation was later proved in a more general setting by Arnold (1986,1998) \cite{Arnold86,Arnold98}.

It has long been known that $H^{M}$ is a ``robust'' invariant since it is conserved
even with a finite but very small Spitzer resistivity as was conjectured by  Taylor (1974) \cite{Taylor74} and proved by Berger (1984) \cite{Berger84}. This is important in explaining the observed tendency of magnetic
fields to evolve towards a force-free configuration in a myriad of situations. This so-called Taylor-Woltjer relaxation  \cite {Woltjer58a,Taylor74} hinges on the conservation of $H^{M}$ as the system loses energy.

It has been argued using mean field theory \cite{Steenbecetal66}, turbulence closure \cite{Frischetal75}, and physical models \cite{Blackman05} that $H^{M}$ undergoes an inverse cascade 
to larger scales rather than being transferred to small resistive scales. This would imply that, unlike in the case 
for energy, turbulence cannot catalyze the dissipation of magnetic helicity.
While the inverse cascade of $H^{M}$ is widely accepted, the possibility of a concurrent
forward cascade of magnetic helicity to small-scales has been raised by 
\cite{Alexakisetal06,Alexakisetal07} based on the analysis of numerical simulations.

In this section, we will present a rigorous proof showing that under
very weak conditions, it is impossible for $H^{M}$ to cascade to arbitrarily small scales. 
Therefore, the direct cascade of magnetic helicity observed from numerical simulations 
cannot persist with increasing magnetic Reynolds numbers.

The Magnetic Helicity balance for the large-scales is: 
\begin{eqnarray} 
\partial_t(\fA\cdot\fB) &+& \grad\cdot\left[  c\,\OL{\bE} \btimes\OL{\bA} + c \,\OL{\phi}~ \OL{\bB} \right] 
=2c\,\bepsilon\bdot\fB -2\eta\frac{4\pi}{c}\ol{\bJ}\cdot\fB 
\lb{large-HMII}\end{eqnarray} 
where $\bB=\grad\btimes\bA$ and $\phi $ is an electrostatic potential satisfying $\frac{1}{c}\partial_{t}\bA = -\bE - \grad\phi$. 
The filtered electric field inside the divergence is given by eq. (\ref{eq:FltrOhmsLaw}), $\OL\bE_\ell= \fJ_\ell/\sigma -\frac{1}{c}\fu_\ell\btimes\fB_\ell -\bepsilon_\ell$, and contains contributions from the turbulent EMF.
It is worthwhile to remark that  conservation of $H^M$ is valid for any velocity field, including that of compressible flows. In fact, eq. (\ref{large-HMII}) and its unfiltered version do not require that velocity $\bu$ is a solution of the momentum equation.

Magnetic helicity defined as $\int_{V}\bA\cdot\bB ~d^{3}\bx$ is gauge-dependent and, thus, 
is ill-defined if the surface of $V$ is not a magnetic surface satisfying $\bB\cdot\hat{\bn}\big|_{\partial V}\ne 0$. A more appropriate quantity to consider is \emph{relative helicity} \cite{Berger-Field,Finn-Antonsen}:
\be
\Delta H \equiv (\bA+\bA^{*})\cdot (\bB - \bB^{*}),
\lb{eq:RelativeHelicity}\ee
which measures the  helicity relative to a reference field $\bB^{*} = \grad\btimes\bA^{*}$ that extends beyond $V$ while satisfying 
$\bB^{*}\cdot\hat{\bn}\big|_{\partial V} = \bB\cdot\hat{\bn}\big|_{\partial V}$. Since $\Delta H$ is gauge invariant for any $\bB^{*}$ satisfying the aforementioned boundary conditions,
 it is most convenient to choose the reference field to be potential: $\bB^{*}=\bP = \grad\psi = \grad\btimes\bA_{P}$.
 
The balance equation of the large-scale relative helicity is
\begin{eqnarray} 
\lefteqn{ 
\frac{d}{dt}\OL{\Delta H} = \int \partial_{t} \left[ (\fA+\ol{\bA_{P}})\cdot (\fB - \ol{\bP}) \right] ~d^{3}\bx     
}\nonumber \\
& & = 2c\int_{V} \bepsilon\cdot\fB -\eta\frac{4\pi}{c^2}\ol{\bJ}\cdot\fB ~ d^{3}\bx  \nonumber\\
& & +2\oint_{\partial V}  \left[ (\ol{\bA_{P}} \cdot \fu)\fB - (\ol{\bA_{P}}\cdot \fB )\fu - c\,\ol{\bA_{P}} \btimes \bepsilon - \eta \frac{4\pi}{c} \ol{\bJ}\btimes\ol{A}_{P}\right] \cdot \hat{\bf{n}} ~dS
\lb{large-HM}\end{eqnarray}
with the gauge choice of $ \grad\cdot\bA_{P}=0 $ and $ \bA_{P} \cdot \hat{\bf{n}} \big|_{\partial V} = 0 $ to simplify the algebra since $\OL{\Delta H} $ is gauge invariant \cite{Berger88}.

The surface integral is the space-transport of magnetic helicity: the first term,  $(\ol{\bA_{P}} \cdot \fu)\fB$, represents transport due to Alfv\'en waves such as torsional waves arising from the twisting motions on the surface and
propagating along large-scale magnetic field lines \cite{Berger99}. The second term is advection of magnetic helicity by the flow. The third term is magnetic helicity's analogue of the turbulent Poynting flux due to the effective electric field $\bepsilon$ resulting from small-scale fluctuations. The last term inside the surface integral 
is diffusive transport due to Spitzer resistivity.

The volume integral on the RHS is the rate of change of magnetic helicity in the volume. 
The second term, $\eta\frac{4\pi}{c^2}\ol{\bJ}\cdot\fB$ is resistive destruction (or creation) of large-scale magnetic helicity. Following a proof similar to that in Proposition \ref{Prop2}, it can be shown to vanish in the limit of $\eta\to 0$.
The first term, $2c\,\fB\cdot\bepsilon$, describes the generation of knottedness in the large scale magnetic field lines due to an effective electric field arising from the small-scales. Since the unfiltered $\Delta H$ is an invariant, this term is therefore a flux of magnetic helicity across scales.
This mechanism is sketched in Figure \ref{FigHelicity} in which
the small scales give rise to the \emph{turbulent EMF} $\bepsilon$ along large-scale magnetic loops, which acts as an electric field to induce a large-scale $\fB$ through the loop. This generates flux linkage between closed $\fB$ lines. However, the flux term $2c\,\fB\cdot\bepsilon$ is not sign definite and can just as well transfer magnetic helicity from large to small scales. The following theorem shows that the latter scenario cannot be sustained to arbitrarily small scales and, therefore, a forward (downscale) cascade of magnetic helicity is not possible under very weak assumptions. Specifically, we assume that 3rd-order moments of $\bu$ and $\bB$ remain bounded with increasing Reynolds numbers. These assumptions are {\it almost} as weak as the condition that kinetic and magnetic energy remain finite with increasing Reynolds numbers. The following theorem was derived in collaboration with Gregory L. Eyink.

\begin{figure}
\centering
\includegraphics[totalheight=.3\textheight,width=.7\textwidth]{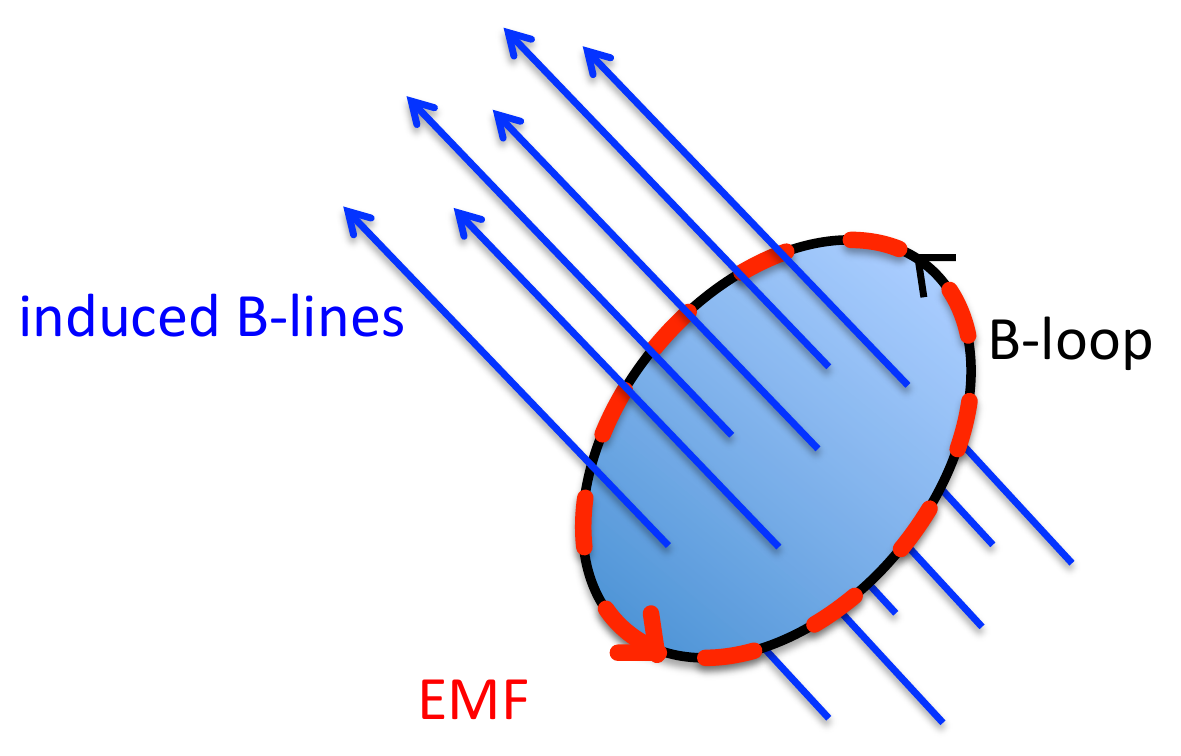}
\caption{Generating large-scale magnetic helicity. Small scales give rise to the turbulent EMF $\bepsilon$ along large-scale magnetic loops. The EMF acts as an electric field to induce a large-scale magnetic field through the loop. This generates flux linkage between closed $\fB$ lines and, therefore, produce large-scale magnetic helicity.}
 \lb{FigHelicity}
\end{figure}

\begin{Th}
Let $\bu$ and $\bB$ be a weak solution of the ideal MHD equations over domain $\Omega=\mathbb{T}^3$ or ${\Omega=\mathbb R}^3$. If third-order moments of the fields are finite, 
\begin{eqnarray}
\lefteqn{
\mbox{ If\hspace{.5cm}} \left\langle|\bu|^{3}\right\rangle < \infty 
\mbox{\hspace{.5cm}and\hspace{.5cm}}
 \left\langle|\bB|^{3}\right\rangle < \infty} \nonumber \\ 
& & \mbox{then\hspace{.5cm}}  \lim_{\ell\to 0} ~\left| \left\langle \fB_{\ell}\cdot\bepsilon_{\ell} \right\rangle \right | = 0. 
\nonumber \end{eqnarray}
\lb{Th-HM}\end{Th}
{\it Proof of Theorem \ref{Th-HM}:}\\
We first prove a standard result in functional analysis, that $L_{3}$-norms of increments $\|\delta\bu(\bx;\br)\|_{3}$,  $\|\delta\bB(\bx;\br)\|_{3}$ (or 3rd-order structure functions) are continuous in $\br$ (for example, see \cite{LiebLoss01}).
Since smooth functions are dense in $L_3(\Omega)$ \cite{LiebLoss01}, $\bu(\bx)$ and $\bB(\bx)$ can be approximated by smooth fields $\bu^{*}(\bx)$ and $\bB^{*}(\bx)$ in $C^{\infty}(\Omega)$, respectively, such that
$\|\bu(\bx)-\bu^{*}(\bx) \|_{3} < \epsilon $ and $\|\bB(\bx)-\bB^{*}(\bx) \|_{3} < \epsilon $ for any positive $\epsilon$. 
It follows that
\begin{eqnarray} 
\hspace{-2cm}
\| \delta\bu(\bx;\br)\|_{3} 
&=& \| \bu(\bx+\br) - \bu(x)\|_{3} \nonumber \\
&\le& \|\bu(\bx+\br) - \bu^{*}(\bx+\br) \|_{3} + \|  \bu^{*}(\bx) - \bu(\bx) \|_{3} + \|  \bu^{*}(\bx + \br) - \bu^{*}(\bx) \|_{3}  \nonumber \\
&\le& 2\epsilon + \|\delta\bu^{*}(\bx;\br)\|_{3} \nonumber
\end{eqnarray}
Hence, 
\begin{eqnarray}  
\limsup_{r\to 0} \| \delta\bu(\bx;\br)\|_{3} \le  2\epsilon + \limsup_{r\to 0} \|\delta\bu^{*}(\bx;\br)\|_{3},
\nonumber\end{eqnarray}  
but $\limsup_{r\to 0} \|\delta\bu^{*}(\bx;\br)\|_{3} = \lim_{r\to 0} \|\delta\bu^{*}(\bx;\br)\|_{3}=0$ by continuity of $\bu^{*}$. Therefore, $L_3$-norms of increments are continuous: $\lim_{r\to 0} \| \delta\bu(\bx;\br)\|_{3}=0$.

Using this standard result, the proof of the theorem follows from an application of the H\"older inequality:
\begin{eqnarray} 
\hspace{-2cm}
\left| \left\langle \fB_{\ell}\cdot\bepsilon_{\ell} \right\rangle \right |
&\le& \left\|\fB\cdot\bepsilon\right\|_{1} \nonumber \\
&\le& \left\|\fB\right\|_{3} \left\|\bepsilon\right\|_{\frac{3}{2}} \nonumber \\
&\le& \left\|\fB\right\|_{3} \bigg( \langle \|\delta\bu(\bx;\br)\|_{3}   \|\delta\bB(\bx;\br)\|_{3}\rangle_{\ell}  +  \langle \|\delta\bu(\bx;\br)\|_{3} \rangle_{\ell} \langle \|\delta\bB(\bx;\br)\|_{3}\rangle_{\ell} \bigg),  \nonumber \\
 \nonumber \end{eqnarray}
The last line follows from identity (\ref{EMF-identity}), with the averaging done over $|\br| < \ell$.
By continuity of $L_{3}$-norms of increments, the terms in parentheses vanish with $\ell\to 0$. 
We are able to prove Theorem \ref{Th-HM} under conditions much weaker compared to those of 
Theorem \ref{Th-Energy} because of the lack of a derivative in the magnetic helicity flux.

Theorem \ref{Th-HM} shows that magnetic helicity is a very robust invariant, requiring infinite-third order moments $\langle |\bu|^{3}\rangle$ and/or $ \langle |\bB|^{3} \rangle $, for 
the turbulent plasma to dissipate magnetic helicity by a non-linear cascade to small scales. 
The result is an improvement over Taylor (1974) \cite{Taylor74} and Berger (1984)\cite{Berger84} because it does not depend on the specifics of microscopic non-idealities. This is also a significant improvement over Theorem 4.2 in \cite{Caflisch97}.

Our result is relevant in the limit of large magnetic Reynolds number and holds for any magnetic Prandtl number, including the limit of large  $Pm$ that exist in many astrophysical systems.  
Moreover, our result also holds for compressible flows since budget (\ref{large-HM}) holds for any velocity and Theorem \ref{Th-HM} assumed only finite 3rd order moments of the velocity.

\section{Cross-Helicity\lb{sec:CrossHelicity}}
Woltjer in 1958 \cite{Woltjer58b} discovered a third quadratic invariant, cross-helicity, $H^{C}$, which measures the degree of mutual knottedness of magnetic field lines with 
vorticity lines. Conservation of $H^{C}$ can be viewed as resulting from the conservation of circulation $\oint_{C(t)}\bu\cdot d\bx$, otherwise known as Kelvin's Theorem, along closed loops of magnetic 
field lines $C(t)$. 
This is because the Lorentz force $\bJ\btimes\bB$ along magnetic field lines is zero.
Cross-helicity is also dynamically relevant
because it measures the alignment of $\bu$ with $\bB$ \cite{Dobrowolny80,Grappin86,Pouquetetal86}, which suppresses the time evolution of $\bB$, as can be seen from the 
induction equation (\ref{B-eq}). Yet, a total shut-down of the induction term is not expected, in general, since the Lorentz force $\bJ\btimes\bB$ tends to create velocities
 perpendicular to $\bB$, unless $\bB$ is a force-free field.

Boldyrev's phenomenological theory of turbulence \cite{Boldyrev05,Boldyrev06} is  based on the joint cascade of energy and cross-helicity. The theory posits that the
cascade of energy to small scales is stronger than the cascade of cross-helicity, thus prohibiting a perfect alignment of small scale velocity magnetic fields. This section presents rigorous constraints on the 
flux of cross-helicity to smaller scales, similar to what was done for energy and magnetic helicity.

The large-scale cross helicity balance is:
\begin{eqnarray}
\hspace{-2cm}\partial_t(\fu\cdot\fB) &+& \grad\cdot \Bigg[ \left(\frac{1}{\rho}\ol{P}  - \frac{|\OL\bu|^2}{2}\right)\OL\bB + (\fu\cdot\fB)\fu + c\,\fu\btimes\bepsilon + \OL\bB\bdot\frac{\btau_\ell}{\rho} + \OL\bB\left(\frac{\ol{\left|\bB\right|^2}}{8\pi\rho} - \frac{\left|\fB\right|^{2}}{8\pi\rho}\right) \nonumber\\
&& \hspace{1cm}-\nu(\grad\OL\bu)\bdot\OL\bB-\eta(\grad\OL\bB)\bdot\OL\bu\Bigg]\nonumber\\
&=& -\Pi^{H^C}_{\ell} -(\nu+\eta)\grad\OL\bB:\grad\OL\bu
\lb{HC-ideal-ell}\end{eqnarray}
The terms inside the divergence represent space transport of large-scale cross-helicity: 
the first term is due to large-scale pressure gradients, the second is due to advection by the large-scale flow, 
the third and fourth terms are due to turbulent diffusion arising from the turbulent EMF and sub-scale kinetic energy, and the last two terms are due to diffusion by microphysical processes. The second term on the RHS is microphysical destruction (or creation) of large-scale cross-helicity. Following a proof similar to that in Proposition \ref{Prop2}, it can be shown to vanish at every point $\bx$ in the limit of $\nu,\eta\to 0$.
The flux term on the RHS is defined as
\be
\Pi^{H^C}_{\ell}(\bx) \equiv -\frac{1}{\rho}\grad\OL\bB_\ell{\bf :} \btau_\ell - c\,\fomega_\ell\cdot\bepsilon_\ell~,
\lb{HC-flux}\ee
where $\bomega=\grad\btimes\bu$ is vorticity. The flux can be rewritten as 
\be
\Pi^{H^C}_{\ell}(\bx) \equiv -\frac{1}{\rho}(\bF^{u}_\ell+\bF^{B}_\ell)\cdot\fB - c\,\fomega_\ell\cdot\bepsilon_\ell 
- \grad\bdot\left[\frac{1}{\rho}\OL\bB\bdot\left(\btau_\ell - \frac{1}{3}{\bf I}\frac{\mbox{trace}\left(\btau_\ell\right)}{2}\right)\right].
\lb{HC-flux_exp2}\ee
Here, ${\bf I}$ is the identity rank-2 tensor. The \emph{turbulent vortex force} \cite{Eyink2006b} $\bF^{u}_\ell$, and the \emph{turbulent Lorentz force} $\bF^{B}_\ell$ in (\ref{HC-flux}) are defined as
\be 
f^{u}_i \equiv \rho (\ol{\bu\btimes\bomega} -\fu\btimes\fomega)_i = -\rho[\partial_j \tau(u_i,u_j) - \frac{1}{2}\partial_i \tau(u_j,u_j)]
\lb{t-vortex-force}\ee
\be 
f^{B}_i \equiv \frac{1}{c} (\ol{\bJ\btimes\bB} -\fJ\btimes\fB)_i = \frac{1}{4\pi}[\partial_j\tau(B_i,B_j) - \frac{1}{2}\partial_i \tau(B_j,B_j)]
\lb{t-Lorentz-force}\ee
It is worthwhile to remark that even though the Lorentz
force $\bJ\btimes\bB$ is always perpendicular to $\bB$ (unless $\bJ\btimes\bB=\bzed$), the \emph{turbulent Lorentz force} $\bF^{B}$ 
can have a component parallel to either $\fB$ or $\bB$. The same is true for $\bepsilon$ which can have a component
 parallel to either $\fu$ or $\bu$ unlike the electric field in the ``bare'' Ohm's law.

From expression (\ref{HC-flux_exp2}), one observes that the mechanism by which $\Pi^{H^C}_{\ell}$ generates (or destroys) large-scale cross-helicity is similar to that of magnetic helicity sketched in Figure \ref{FigHelicity}. Note that the third term, being a divergence, vanishes after averaging over space. The turbulent forces $\bF^{u}_\ell+\bF^{B}_\ell$ accelerate the flow along a large-scale magnetic loop thus creating a flux of vorticity through the loop. The vortex lines threading the loop are closed due to the solenoidal nature of $\fomega$, which creates knotted large-scale vortex and magnetic field lines. The same turbulent forces can just as well destroy large-scale cross-helicity by decelerating the flow along a large-scale magnetic loop. Similarly, the turbulent EMF $\bepsilon$ along vortex loops $\fomega$ induces a magnetic flux through the $\fomega$-loops.
Since the turbulent forces result from the fluctuations at scales smaller than $\ell$, and since $H^C$ is a conserved quantity, such a mechanism is necessarily a flux of $H^{C}$ across scales.

The conditions required for the turbulent plasma is to sustain a cascade of cross-helicity 
to arbitrarily small scales are qualitatively similar to those required to cascade energy. 
The proof is similar to that of Theorem \ref{Th-Energy}, whereby 
 we first express the cross-helicity flux $\Pi^{H^C}_{\ell}$ in terms of increments:
\begin{eqnarray}
\hspace{-2cm}\Pi^{H^C}_{\ell}(\bx) 
&\equiv& -\frac{1}{\rho}\grad\OL\bB_\ell{\bf :} \btau_\ell - \fomega_\ell\cdot\bepsilon_\ell\nonumber\\
&=& O\left(\frac{\delta B(\ell)}{\ell} \cdot \left(\delta u^2(\ell)+\frac{\delta B^2(\ell)}{4\pi\rho}\right)\right)
+O\left(\frac{\delta u(\ell)}{\ell} \cdot \delta u(\ell)\delta B(\ell)\right)
\end{eqnarray}
Therefore, the flux scales as
\be
\Pi^{H^C}_{\ell} \sim  {\mathcal O}\left(\min\left\{\ell^{3\sigma^b-1},\ell^{2\sigma^u+\sigma^b-1}\right\}\right).
\ee
If $\sigma^b>1/3$ and $2\sigma^u+\sigma^b>1$, then $\Pi^{H^C}_{\ell}$ will decay as $\ell\to 0$. In other words, if the nonlinearities in MHD are to cascade cross-helicity to arbitrarily small scales, either $\sigma^b\le1/3$ or $2\sigma^u+\sigma^b\le1$ (the 'or' is non-exclusive). The following theorem, which was derived in collaboration with Gregory L. Eyink,
shows this rigorously.

\begin{Th} 
Let $\bu$ and $\bB$ be a weak solution of the ideal MHD equations over domain $\Omega={\mathbb T}^3$ or $\Omega={\mathbb R}^3$. Assume $\| \delta_{r}\bu\|_{3}  \sim r^{\sigma_{3}^{u}}$  and $\| \delta_{r}\bB\|_{3}  \sim r^{\sigma_{3}^{b}}$.
\begin{eqnarray}
\lefteqn{
\mbox{If\hspace{.5cm}} \sigma_{3}^{b} > \frac{1}{3} \mbox{\hspace{.5cm}and\hspace{.5cm}} \sigma_{3}^{b}+ 2\sigma_{3}^{u} > 1,
} \nonumber \\ 
& & \mbox{then\hspace{.5cm}}  \lim_{\ell\to 0} ~\left | \left\langle \Pi^{H^C}_{\ell}  \right\rangle\right |
\le \lim_{\ell\to 0}  \left( \const~ \ell^{3\sigma_{3}^{b}-1} + \const~ \ell^{2\sigma_{3}^{u}+\sigma_{3}^{b}-1}  \right)
= 0. 
\nonumber \end{eqnarray}
\lb{Th-HC}\end{Th}
{\it Proof of Theorem \ref{Th-HC}:}\\
The proof follows from an application of the H\"older inequality:
\begin{eqnarray} 
\hspace{-2.5cm}\left | \left\langle \Pi^{H^C}_{\ell}  \right\rangle\right | 
&\le& \|\grad\fB_\ell:\btau_\ell \|_{1} + \|\fomega_\ell\cdot\bepsilon_\ell \|_{1}  \nonumber\\
&\le& \|\grad\fB_\ell \|_{3} \|\btau_\ell \|_{\frac{3}{2}} + \|\fomega_\ell \|_{3} \|\bepsilon_\ell \|_{\frac{3}{2}}  \nonumber\\
&\le& ~~~\bigg\| \frac{1}{\ell} \int d^{3}r ~ (\grad G)_{\ell}(r) ~ \delta_{r}\bB(\bx)  \bigg\|_{3} ~~  
 \bigg \| \big(\langle\delta_{r}\bu\delta_{r}\bu\rangle_\ell-\langle \delta_{r}\bu\rangle_\ell\langle \delta_{r}\bu\rangle_\ell \big)\bigg \|_{\frac{3}{2}}\nonumber\\
&&+ \bigg\| \frac{1}{\ell} \int d^{3}r ~ (\grad G)_{\ell}(r) ~ \delta_{r}\bB(\bx)  \bigg\|_{3} ~~ 
\bigg \|\big(\langle\delta_{r}\bB\delta_{r}\bB\rangle_\ell-\langle\delta_{r}\bB\rangle_\ell\langle\delta_{r}\bB\rangle_\ell \big) \bigg\|_{\frac{3}{2}}  
 \nonumber\\
&&+ \bigg\| \frac{1}{\ell} \int d^{3}r ~ (\grad G)_{\ell}(r) \btimes \delta_{r}\bu(\bx) \bigg\|_{3}  ~~    \bigg\| \langle \delta_{r} \bu\btimes\delta_{r}\bB \rangle-\langle \delta_{r} \bu\rangle\btimes\langle\delta_{r}\bB \rangle \bigg\|_{\frac{3}{2}}  ~~~~~~~~~~~~    \nonumber\\
&\le& \const\ell^{\sigma_{3}^{b}-1} \left(\const\ell^{2\sigma_{3}^{u}} + \const\ell^{2\sigma_{3}^{b}}\right) + \const\ell^{\sigma_{3}^{u}-1}\left(\const\ell^{\sigma_{3}^{u}+\sigma_{3}^{b}}\right)  \nonumber\\
&=& \const\ell^{2\sigma_{3}^{u}+\sigma_{3}^{b}-1} + \const\ell^{3\sigma_{3}^{b}-1}.  
\nonumber \end{eqnarray}
The upper bound vanishes in the limit of $\ell\to 0$, thus proving our result.

\section{Summary\lb{sec:Summary}}

In this paper, we formulated a coarse-graining approach to analyze the fully nonlinear dynamics of 
MHD plasmas and liquid metals. Using this methodology, we derived effective equations for the observable 
velocity and magnetic fields spatially-averaged at an arbitrary scale of resolution. 
These macroscopic effective equations contain both a ``subscale stress'' and a ``subscale  EMF'' 
generated by nonlinear interaction of eliminated plasma motions.
Despite its close resemblance to mean-field MHD, commonly employed in dynamo theory \cite{KrauseRaedler80, Biskamp03},
the ``coarse-graining'' approach allows for the description of dynamics at any arbitrary scale. 
Furthermore, such a description is deterministic, valid at every point in space-time, without requiring any statistical averaging or any assumption of scale separation. 

Using this scale-decomposition framework, we proved rigorously 
that the direct role of molecular viscosity and Spitzer resistivity in 
the evolution of the large-scales is negligible at every point in space
and at all times. The evolution of the large scales can be influenced instead 
by nonlinear effects from smaller scales. These small-scales exert stresses 
(both Reynolds and Maxwell stresses) as well as generate electric fields which
can play a major role in the evolution of the large scales.

We then established local balance equations in space-time of the three quadratic invariants 
--energy, cross helicity, and magnetic helicity-- for measurable ``coarse-grained'' variables.
Particular attention was given to the effects of sub-scale terms accounting for the modes eliminated from
the coarse-grained dynamical equations. Physical interpretations of these terms, which are responsible for 
the turbulent cascades in MHD flows, were presented in terms of work concepts for energy and in terms of 
topological flux-linkage \cite{Moffatt69} for the two helicities. The subscale nonlinear terms also contribute to 
enhanced spatial transport of the MHD invariants, which dominate over microphysical transport at large-scales.  

We derived rigorous constraints on the cascade of these quadratic invariants. 
In order for the nonlinear terms to sustain a cascade of energy \cite{Caflisch97} 
and cross-helicity to arbitrarily small scales, it is necessary that the velocity and 
magnetic fields be rough enough. This roughness is reflected in the scaling 
exponents of structure functions. 

We also proved that the conditions required 
for magnetic-helicity to undergo a forward cascade to arbitrarily small scales
are almost as severe as requiring infinite energy. 
We emphasize that our result does not preclude the transfer of a finite
amount of magnetic helicity to resistive scales at relatively moderate
magnetic Reynolds numbers, such as in the case of numerical simulations that are feasible with 
today's computational resources.
However, our result proves that such transfer to the dissipation scales will vanish
with increasing magnetic Reynolds number. In general, when analyzing simulations of 
turbulent flows, it is vitally important to study trends as a function of Reynolds number
and check if the phenomenon under study persists and can be extrapolated to the 
large Reynolds numbers present in nature. 

The results of this paper lay out rigorous constraints which have to be satisfied by 
any phenomenological theory of MHD turbulence.

\subsection{Acknowledgments}
The mathematical proofs included here were derived in collaboration with 
Gregory L. Eyink, who also contributed to the content of this paper. I also thank 
E. T. Vishniac for valuable discussions, and three anonymous referees for useful 
comments and suggestions. This work was supported in part by the DOE Office 
of Fusion Energy Sciences grant DE-SC0014318, by the DOE National Nuclear 
Security Administration under award DE-NA0001944, by NSF grant OCE-1259794, 
and by the LANL LDRD program through project number 20150568ER.\\

\end{document}